\documentclass{article}
\usepackage[linesnumbered]{algorithm2e}
\DontPrintSemicolon
\usepackage{multirow}
\usepackage{tikz}
\usepackage{PRIMEarxiv}
\usepackage[sorting=none]{biblatex}
\bibliography{references}
\usepackage[utf8]{inputenc}
\usepackage[T1]{fontenc}    
\usepackage{hyperref}       
\usepackage{url}            
\usepackage{booktabs}       
\usepackage{amsfonts}       
\usepackage{nicefrac}       
\usepackage{microtype}      
\usepackage{lipsum}
\usepackage{fancyhdr}       
\usepackage{graphicx}       
\usepackage{amsmath} 
\usepackage{multicol}
\usepackage{caption}
\graphicspath{{media/}}     
\usepackage{todonotes}

\RestyleAlgo{ruled} 

\title{Control-Informed Reinforcement Learning for Chemical Processes}

\author{
  Maximilian Bloor, Akhil Ahmed, Niki Kotecha, \\ \textbf{Mehmet Mercangöz, Calvin Tsay\thanks{Corresponding Authors}, Ehecactl Antonio Del Rio Chanona$^*$} \\
  Sargent Centre for Process Systems Engineering, Department of Chemical Engineering\\
  Imperial College London \\
  London\\
  \texttt{\{max.bloor22, a.del-rio-chanona, c.tsay\}@imperial.ac.uk} }

\usetikzlibrary{shapes.geometric, shapes, arrows, positioning, calc}
\begin{document}
\captionsetup[table]{hypcap=false}
\captionsetup[figure]{hypcap=false}
\maketitle

\begin{abstract}
This work proposes a control-informed reinforcement learning (CIRL) framework that integrates proportional-integral-derivative (PID) control components into the architecture of deep reinforcement learning (RL) policies. The proposed approach augments deep RL agents with a PID controller layer, incorporating prior knowledge from control theory into the learning process. CIRL improves performance and robustness by combining the best of both worlds: the disturbance-rejection and setpoint-tracking capabilities of PID control and the nonlinear modeling capacity of deep RL.
Simulation studies conducted on a continuously stirred tank reactor system demonstrate the improved performance of CIRL compared to both conventional model-free deep RL and static PID controllers. CIRL exhibits better setpoint-tracking ability, particularly when generalizing to trajectories outside the training distribution, suggesting enhanced generalization capabilities. Furthermore, the embedded prior control knowledge within the CIRL policy improves its robustness to unobserved system disturbances. The control-informed RL framework combines the strengths of classical control and reinforcement learning to develop sample-efficient and robust deep reinforcement learning algorithms, with potential applications in complex industrial systems.
\end{abstract}

\keywords{Reinforcement learning \and Process control \and PID Control}

\section{Introduction}
In the chemical process industry, maintaining control over complex systems is crucial for achieving reliable, efficient, and high-performance operations~\cite{simkoff2020process}. Traditionally, process control has relied heavily on classical feedback control techniques such as proportional-integral-derivative (PID) controllers due to their simplicity, interpretability, and well-established tuning methods \cite{PIDsurvey}. However, these tuning methods are often largely empirical, or otherwise rely on having accurate mathematical models of the open-loop system dynamics and disturbance responses, which can be challenging to derive for complex processes involving nonlinearities, delays, constraints, and changing operating conditions~\cite{seborg2016process}. 
As a result, PID controllers often struggle to provide adequate control performance without extensive re-tuning or gain scheduling \cite{velocityref}.
One alternative is Model Predictive Control (MPC), a successful model-based process control strategy, with the ability to optimize control actions based on the current system states and predicted future behavior while satisfying constraints. This has led to its widespread adoption in the chemical process industry~\cite{forbes2015model}. Typically, MPC operates as a supervisory layer below the real-time optimization (RTO) layer, providing control setpoints to lower-level regulatory controllers, often PID controllers, which directly manipulate process variables. This hierarchical structure combines the predictive capabilities of MPC with the rapid response of traditional feedback control. However, the performance of MPC heavily relies on the accuracy of its internal process model~\cite{MPCreviews}. The ongoing digitalization of the chemical process industry has opened new avenues for enhancing MPC performance through data-driven approaches. This digital transformation has enabled the integration of advanced analytics and machine learning techniques into both MPC and regulatory control layers, especially on the modeling end. For instance, artificial neural networks~\cite{piche2000nonlinear} and Gaussian processes~\cite{kocijan2004gaussian} have been employed to capture complex process dynamics that may be challenging to model using purely mechanistic approaches. Furthermore, the increased availability of process data has facilitated the development of hybrid models that combine first-principles knowledge with data-driven components~\cite{zhang2019real, sorourifar2021data}. These advancements, coupled with improvements in computational capabilities, have expanded the applicability of MPC to more complex and uncertain processes. However, challenges remain in areas such as online computational requirements for large-scale systems and the handling of uncertainties.

Recently, reinforcement learning (RL) has emerged as a promising data-driven framework for learning control policies directly from interactions with the chemical process system \cite{nian2020review}. Deep RL methods, which utilize deep neural networks, have demonstrated success in a variety of difficult decision-making and control problems. A key advantage of model-free RL is that it does not require accurate system models once online, instead learning control policies from experience. While model-free RL approaches can learn control policies without requiring explicit system models, they often face challenges in sample efficiency and may not fully leverage existing domain knowledge \cite{LAWRENCESensingcontrolreview}.

For safety reasons, RL algorithms in chemical process control are typically trained on simulation models rather than directly on physical systems. Despite this limitation, RL offers several advantages over traditional control methods. One significant benefit is the fast online inference time. This characteristic makes RL particularly suitable for systems where online computation time is critical, as the trained policy can execute control decisions rapidly in real-time applications. RL also shows promise in handling complex, nonlinear systems and adapting to process uncertainties. This feature allows RL to potentially address challenges in dynamic chemical processes more effectively than traditional control approaches. We note there is also growing interest in algorithms for \textit{safe} RL, or those which can avoid (known or unknown) constraints, e.g., in physical systems~\cite{gu2022review}.

It is important to acknowledge the challenges associated with implementing RL to control complex chemical processes. The offline training of RL agents often requires a large number of samples to achieve satisfactory performance, making the training process computationally intensive and time-consuming. The quality of the trained agent is highly dependent on the fidelity of the simulation model used, which may not always capture all the nuances of real-world processes, requiring online fine-tuning. Moreover, deep RL methods often treat the control problem as a black box, failing to incorporate valuable insights from control theory. This highlights the need for approaches that can balance the model-free learning capabilities of RL with the incorporation of domain expertise and efficient exploration strategies.

\subsection{Related works}

Deep reinforcement learning (deep RL), which combines deep neural networks (DNNs) with RL, has been demonstrated in various domains, including robotics, data center operations, and playing games \cite{badgwell2018reinforcement, roboticsRL, Silver2017}. This success has brought attention to RL from the process systems and control communities. The process systems engineering community has made significant progress in adapting RL to the process industries, including in distributed systems \cite{TANG201836}, constraint handling \cite{BURTEA2024108518, PETSAGKOURAKIS202235, YOO2022157}, inventory management \cite{BURTEA2024108518, mousa2023}, batch bioprocess and control \cite{PETSAGKOURAKIS2020106649, YOO2021108, benchmarkbatch}, production scheduling \cite{HUBBS2020106982}, and energy systems \cite{ALABI2023120633}. Early applications of RL in process control proposed model-free RL for tracking control and optimization in fed-batch bioreactors \cite{kaisare2003simulation, wilson1997neuro,peroni2005optimal}. More recently, \textcite{mowbray2021using} employed a two-stage strategy using historical process data to warm-start the RL algorithm and demonstrated this on three setpoint-tracking case studies. \textcite{MACHALEK2021107496} developed an implicit hybrid machine learning model combining physics-based equations with artificial neural networks, demonstrating its application for reinforcement learning in chemical process optimization. \textcite{zhu2020scalable} developed an RL algorithm that improves scalability by reducing the size of the action space, which was demonstrated on a plantwide control problem. However, the sample efficiency of these algorithms remains a key aspect restricting their widespread industrial adoption.

To address the limitations, prior works have explored integrating reinforcement learning with existing control structures, e.g., PID controllers \cite{kumar2021diffloop,lakhani2021stability,sedighizadeh2008adaptive}. Early approaches applied model-free RL to directly tune the gains of PID controllers \cite{lee2006approximate,brujeni2010dynamic} or used model-based RL techniques such as dual heuristic dynamic programming \cite{berger2013neurodynamic}. Other approaches have investigated embedding knowledge of the dynamical system using physics-informed neural networks to act as a surrogate model of the process for offline training of the RL agent \cite{FARIA2024107256}. Efforts have also been made to develop interpretable control structures that maintain transparency while leveraging advanced optimization techniques~\cite{paulson2023tutorial}. \textcite{lawrence2022deep} directly parameterized the RL policy as a PID controller instead of using a deep neural network, allowing the RL agent to improve the controller's performance while leveraging existing PID control hardware. This work demonstrates that industry can utilize actor-critic RL algorithms without the need for additional hardware or the lack of interpretability which often accompanies the use of a deep neural network. To improve this work's training time, \textcite{MCCLEMENT2022139} used a meta-RL approach to tune PI controllers offline. The method aimed to learn a generalized RL agent on a distribution of first-order plus time delay (FOPTD) systems, resulting in an adaptive controller that can be deployed on new systems without any additional training. However, while the meta-RL approach removes the need for explicit system identification, some knowledge of the process gain and time constant magnitudes is still required to appropriately scale the meta-RL agent's inputs and outputs when applying it to new systems.

\subsection{Contributions}
In contrast to the above methods that focus on tuning PID gain values with a fixed control structure, we propose a control-informed reinforcement learning (CIRL) framework that integrates the PID control structure with a deep neural network into the control policy architecture of an RL agent. This allows the approach to adapt to changing operating points due to the inclusion of the deep neural network. Furthermore, it aims to leverage the strengths of both PID control and deep RL: we seek to improve sample efficiency and stability using known PID structures while gaining robustness and generalizability from RL. In summary, the key contributions of this work are as follows:

\begin{enumerate}
    \item We introduce the CIRL framework, which augments deep RL policies with an embedded PID controller layer. This enables the agent to learn adaptive PID gain tuning while preserving the stabilizing properties and interpretability of PID control, effectively acting as an automated gain scheduler.

    \item We demonstrate the CIRL framework on a nonlinear continuously stirred tank reactor (CSTR) system. The CIRL agent improves setpoint tracking performance compared to both a static PID controller and a standard model-free deep RL approach, particularly when generalizing to operating regions outside the training distribution.
    
    \item We show that by leveraging the embedded prior knowledge from the PID structure, the CIRL agent exhibits enhanced robustness to process disturbances that are not observable during training.
\end{enumerate}

The remainder of this article is organized as follows. Section \ref{sec:back} presents the background on PID control and reinforcement learning. Section \ref{sec:method} describes the proposed control-informed reinforcement learning framework in detail. Section \ref{sec:results} discusses the simulation and experimental results. Finally, Section \ref{sec:conclusion} concludes the article and outlines future research directions.

\section{Background}\label{sec:back}

\subsection{Reinforcement Learning}

\begin{figure}[htbp]
    \centering
    \includegraphics[clip, trim=1.1in 3.1in 2.9in 0.5in, width=0.5\textwidth]{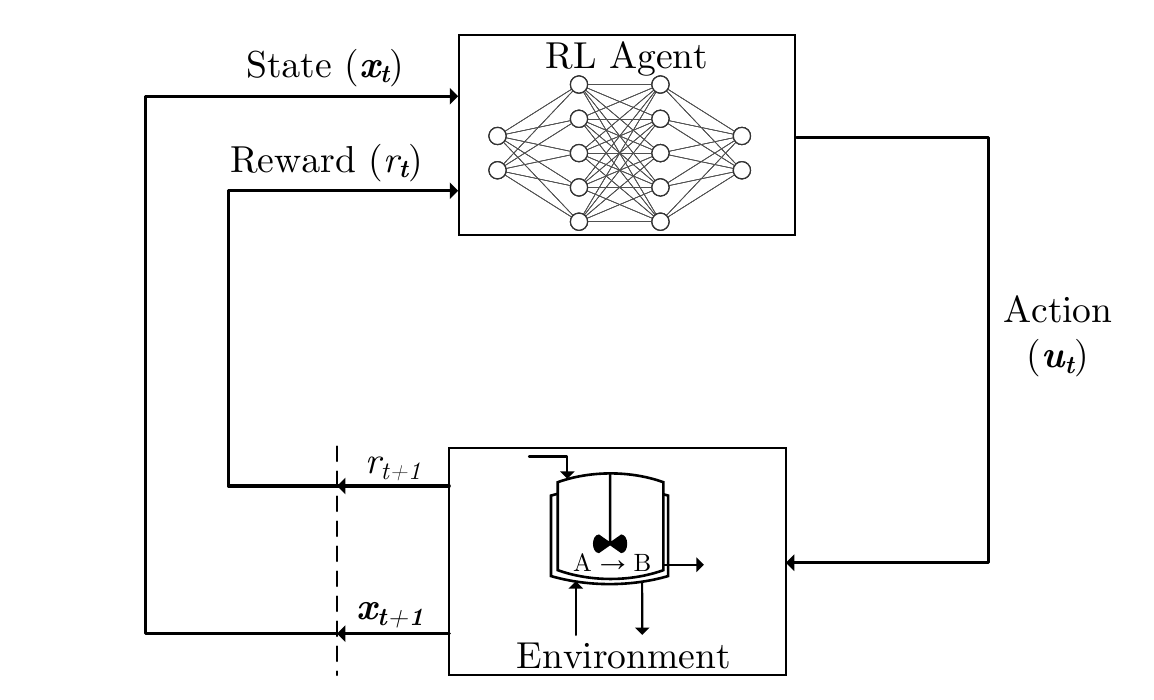}
    \caption{The RL framework}
    \label{fig:RL_diagram}
\end{figure}

The standard RL framework (Figure \ref{fig:RL_diagram}) consists of an agent that interacts with an environment. Assuming the states are fully observable, the agent receives a vector of measured states $\boldsymbol{x}_t \in \mathcal{X} \subseteq \mathbb{R}^{n_x}$, and can then take some action $\boldsymbol{u}_t \in \mathcal{U} \subseteq  \mathbb{R}^{n_u}$, which results in the environment progressing to state $\boldsymbol{x}_{t+1}$. Sets $\mathcal{X}$ and $\mathcal{U}$ represent the state and action space, respectively. 
For a deterministic policy $\pi$, the agent takes actions $\boldsymbol{u}_t = \pi(\boldsymbol{x}_t)$, while, for a stochastic policy, the action $\boldsymbol{u}_t$ is sampled from the policy $\pi$ represented by a conditional probability distribution $\boldsymbol{u}_t \sim \pi(\cdot \mid \boldsymbol{x}_t)$. 
A common assumption in RL is that the state transition given some action is defined by a density function $\boldsymbol{x}_{t+1} \sim p(\cdot \mid \boldsymbol{x}_t,\boldsymbol{u}_t)$ that represents the stochastic nonlinear dynamics of the process. The reward the agent receives is defined by the function $r_t = \mathcal{R}(\boldsymbol{x}_t,\boldsymbol{u}_t)$. With a defined control policy, the policy can be implemented over a discrete time horizon $T$ thus producing the following trajectory $\tau = (\boldsymbol{x}_0, \boldsymbol{u}_0, r_0, \boldsymbol{x}_1, \boldsymbol{u}_1, r_1, ..., \boldsymbol{x}_T,\boldsymbol{u}_T,r_T)$.

Formally, notice that the above state transition assumption enables modeling the underlying system as a Markov Decision Process (MDP); for further treatment of the subject the reader is referred to \textcite{suttonRL}. In reinforcement learning, particularly in our framework, the agent's goal is to maximize the cumulative reward (return) $J(\pi)$ over a pre-defined (often infinite) timespan, a discount factor $\gamma$ is used to reflect the uncertain future and ensure computational tractability. The policy that achieves this is the optimal policy $\pi^*$:

\begin{equation}\label{eq:ret}
    J(\pi) = \mathbb{E}_\pi \left[\sum^T_{t = 0}\gamma^t r_t(\boldsymbol{x}_t,\pi(\boldsymbol{x}_t))\right]
\end{equation}

\begin{equation}
    \pi^* = \text{arg}\max_\pi J(\pi)
\end{equation}
The value function $V^\pi(\boldsymbol{x}_t)$ represents the expected return starting from state $\boldsymbol{x}_t$ and following policy $\pi$ thereafter.
\begin{equation}
    V^{\pi}(\boldsymbol{x}_t) = \mathbb{E}_{\pi}\left[J(\pi) | \boldsymbol{x}_0 = \boldsymbol{x}\right]
\end{equation}

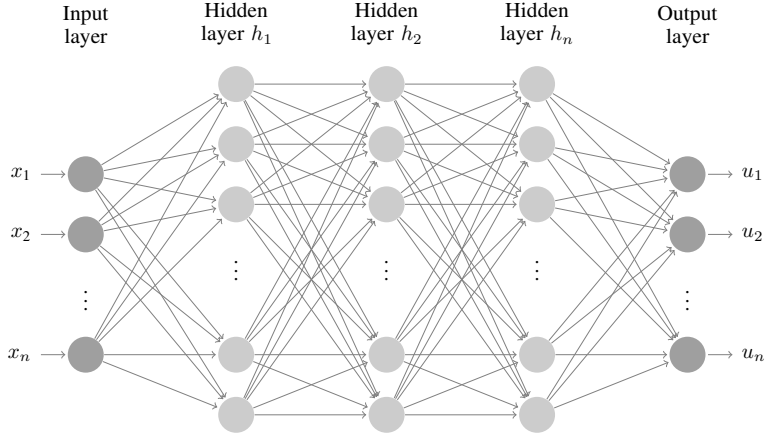
\begin{figure}
\centering
\scalebox{0.8}{
\begin{tikzpicture}[shorten >=1pt,->,draw=black!50, node distance=\layersep]
    \tikzstyle{every pin edge}=[<-,shorten <=1pt]
    \tikzstyle{neuron}=[circle,fill=gray!25,minimum size=17pt,inner sep=0pt]
    \tikzstyle{input neuron}=[neuron, fill=gray!75];
    \tikzstyle{output neuron}=[neuron, fill=gray!75];
    \tikzstyle{hidden neuron}=[neuron, fill=gray!40];
    \tikzstyle{annot} = [text width=4em, text centered]
    \def\numinputs{4}
    \def\numhidden{6}
    \def\numoutputs{4}
    \def\layersep{2.5cm}

    \foreach \name / \y in {1,2,3}
        \path[yshift=0.5cm]
            node[hidden neuron] (H1-\name) at (\layersep,-\y cm) {};

    \node at (\layersep,-3.5) {$\vdots$};

    \node[hidden neuron] (H1-5) at (\layersep,-5 cm) {};
    \node[hidden neuron] (H1-\numhidden) at (\layersep,-6 cm) {};
    \foreach \name / \y in {1,2,3}
        \path[yshift=0.5cm]
            node[hidden neuron] (H2-\name) at (2*\layersep,-\y cm) {};
  
    \node at (2*\layersep,-3.5) {$\vdots$};

    \node[hidden neuron] (H2-5) at (2*\layersep,-5 cm) {};
    \node[hidden neuron] (H2-\numhidden) at (2*\layersep,-6 cm) {};
    \foreach \name / \y in {1,2,3}
        \path[yshift=0.5cm]
            node[hidden neuron] (H3-\name) at (3*\layersep,-\y cm) {};
 
    \node at (3*\layersep,-3.5) {$\vdots$};

    \node[hidden neuron] (H3-5) at (3*\layersep,-5 cm) {};
    \node[hidden neuron] (H3-\numhidden) at (3*\layersep,-6 cm) {};

    \foreach \name / \y in {1,2}
        \node[input neuron, pin=left:$x_\text{\y}$ ] (I-\name) at (0,-\y-1) {};

    \node at (0,-4) {$\vdots$};

    \node[input neuron, pin=left:$x_n$] (I-\numinputs) at (0,-5) {};

    \node[output neuron,pin={[pin edge={->}]right:$u_1$}] (O-1) at (4*\layersep,-2 cm) {};
    \node[output neuron,pin={[pin edge={->}]right:$u_2$}] (O-2) at (4*\layersep,-3 cm) {};
   
    \node at (4*\layersep,-4) {$\vdots$};

    \node[output neuron,pin={[pin edge={->}]right:$u_n$}] (O-\numoutputs) at (4*\layersep,-5) {};

    \foreach \source in {1,2,\numinputs}
        \foreach \dest in {1,2,3,5,\numhidden}
            \path (I-\source) edge (H1-\dest);
    \foreach \source in {1,2,3,5,\numhidden}
        \foreach \dest in {1,2,3,5,\numhidden}
            \path (H1-\source) edge (H2-\dest);
    \foreach \source in {1,2,3,5,\numhidden}
        \foreach \dest in {1,2,3,5,\numhidden}
            \path (H2-\source) edge (H3-\dest);

    \foreach \source in {1,2,3,5,\numhidden}
        \foreach \dest in {1,2,\numoutputs}
            \path (H3-\source) edge (O-\dest);
    \node[annot,above of=H1-1, node distance=1cm] (hl) {Hidden layer $h_1$};
    \node[annot,above of=H2-1, node distance=1cm] (hl) {Hidden layer $h_2$};
    \node[annot,above of=H3-1, node distance=1cm] (hl) {Hidden layer $h_n$};
    \node[annot,above of=I-1, node distance=2.45cm] {Input layer};
    \node[annot,above of=O-1, node distance=2.45cm] {Output layer};
\end{tikzpicture}}
\caption{Deep policy network $\pi_\theta$}
\label{fig:deep-pi}
\end{figure}

Similarly, the action-value function, or ``Q-function,'' $Q^\pi(\boldsymbol{x}_t,\boldsymbol{u}_t)$ represents the expected return from starting from state $\boldsymbol{x}_t$ and taking action $\boldsymbol{u}_t$, assuming that policy $\pi$ is followed otherwise:

\begin{equation}\label{eq:Q}
     Q^{\pi}(\boldsymbol{x}_t,\boldsymbol{u}_t) = \mathbb{E}_{\pi}\left[J(\pi) | \boldsymbol{x}_0 = \boldsymbol{x}, \boldsymbol{u}_0 = \boldsymbol{u}\right]
\end{equation}
Two broad classes of algorithms have emerged to solve the previously described problem: policy optimization methods and value-based methods. These two methods have been effective in deep RL, where a DNN has been used as a function approximator to mitigate the ``curse of dimensionality'' stemming from the discretization of both action and state spaces often necessary to solve continuous problems \cite{sutton1999policy}. The use of DNNs allows the parameterization of the policy $\pi \approx \pi_\theta$ where $\theta \in \Omega \subseteq \mathbb{R}^{n_\theta}$ represents the parameters of the policy and $\Omega$ is the parameter space (Figure \ref{fig:deep-pi}).

A popular group of policy optimization methods leverages policy gradients to optimize the policy in reinforcement learning \cite{Williams1992-ec}. Policy gradient methods, such as Trust Region Policy Optimization (TRPO) \cite{10.5555/3045118.3045319} and REINFORCE \cite{Williams1992-ec}, follow a stochastic gradient ascent strategy to update the policy parameters with a scalar learning rate $\alpha$:
\begin{equation}
    \theta \leftarrow \theta + \alpha \nabla_\theta \widehat{J}(\boldsymbol{\theta})
\end{equation}
These methods directly optimize the expected return $J(\boldsymbol{\theta})$ by following the gradient of the policy parameters $\boldsymbol{\theta}$. The gradient is estimated from sampled trajectories collected by rolling out the current policy in the environment. Policy gradient methods offer several advantages, including their ability to handle continuous action spaces effectively, and the direct optimization of the policy. However, they often suffer from high variance in gradient estimates leading to these methods converging to locally optimal policies. The specific algorithms within this family have their own characteristics; for instance, TRPO~\cite{10.5555/3045118.3045319} provides more stable updates but can be computationally expensive, and REINFORCE~\cite{Williams1992-ec}, while conceptually straightforward, often suffers from high variance in practice.

\begin{figure}
\centering
\scalebox{0.8}{
\begin{tikzpicture}[
    shorten >=1pt,
    ->,
    draw=black!50, 
    node distance=2.5cm,
    neuron/.style={circle,fill=gray!25,minimum size=17pt,inner sep=0pt},
    input neuron/.style={neuron,fill=gray!75},
    output neuron/.style={neuron,fill=gray!75},
    hidden neuron/.style={neuron,fill=gray!40},
    annot/.style={text width=4em, text centered}
]
    \def\ysep{1.7cm}
    \def\inputsep{0.5cm}

    \foreach \i in {1,2}
        \node[input neuron] (Ix-\i) at (0,-\i*1.5*\inputsep+1.5*\ysep-15) {};
    \node at (0,-4*\inputsep+1.2*\ysep) {$\vdots$};
    \node[input neuron] (Ix-3) at (0,-5.5*\inputsep+1.2*\ysep) {};

    \foreach \i in {1,2}
        \node[input neuron] (Iu-\i) at (0,-\i*1.5*\inputsep-5*\inputsep+0.9*\ysep) {};
    \node at (0,-10.1*\inputsep+1.2*\ysep) {$\vdots$};
    \node[input neuron] (Iu-3) at (0,-12*\inputsep+1.2*\ysep) {};

    \foreach \l [count=\x from 1] in {1,2,3}
    {
        \foreach \i in {1,2,3}
            \node[hidden neuron] (H\l-\i) at (\x*2.5,-\i*\ysep+2.25*\ysep) {};
        \node at (\x*2.5,-4*\ysep+2.25*\ysep) {$\vdots$};
        \node[hidden neuron] (H\l-4) at (\x*2.5,-5*\ysep+2.25*\ysep) {};
    }

    \node[output neuron] (O-1) at (10,-3*\ysep+2.25*\ysep) {};

    \foreach \l [remember=\l as \lastl (initially 1)] in {2,3}
        \foreach \i in {1,2,3,4}
            \foreach \j in {1,2,3,4}
                \draw[->] (H\lastl-\j) -- (H\l-\i);

    \foreach \i in {1,2,3}
        \foreach \j in {1,2,3,4}
            \draw[->] (Ix-\i) -- (H1-\j);

    \foreach \i in {1,2,3}
        \foreach \j in {1,2,3,4}
            \draw[->] (Iu-\i) -- (H1-\j);

    \foreach \i in {1,2,3,4}
        \draw[->] (H3-\i) -- (O-1);

    \node[annot,above of=Ix-1, node distance=1.88cm] {Input layer};
    \foreach \l [count=\x from 1] in {1,2,3}
        \node[annot,above of=H\l-1, node distance=1cm] {Hidden layer $h_\l$};
    \node[annot,above of=O-1, node distance=4.4cm] {Output layer $Q$};

    \foreach \i in {1,2}
        \node[left of=Ix-\i, node distance=0.8cm] {$x_\i$};
    \node[left of=Ix-3, node distance=0.8cm] {$x_n$};
    
    \foreach \i in {1,2}
        \node[left of=Iu-\i, node distance=0.8cm] {$u_\i$};
    \node[left of=Iu-3, node distance=0.8cm] {$u_n$};

    \node[right of=O-1, node distance=0.8cm] {$Q$};
\end{tikzpicture}}
\caption{Deep Q-function $Q_\phi$}
\label{fig:deepQ}
\end{figure}
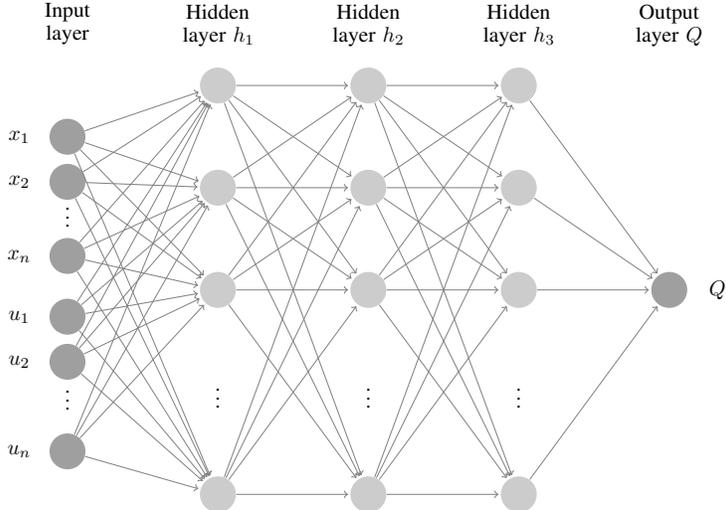

The second class of reinforcement learning methods comprises value-based algorithms, such as Deep Q-Network (DQN) \cite{mnih2013playing}, which rely on learning an action-value (or Q) function. This Q-function can be approximated with a deep neural network with parameters $\phi \in \Phi \subseteq \mathbb{R}^{n_\phi}$ resulting in $Q_\phi$ (Figure \ref{fig:deepQ}). 
The parameters are then updated by minimizing mean squared error against targets given by the Bellman recursion equation as follows:
\begin{equation}
       \mathcal{L}(\phi) = \mathbb{E}_{(\boldsymbol{x}_t,\boldsymbol{u}_t,r_t,\boldsymbol{x}_{t+1})}\left[(r_t + \gamma \max_{\boldsymbol{u}_{t+1}}Q_\phi(\boldsymbol{x}_{t+1},\boldsymbol{u}_{t+1}) - Q_\phi(\boldsymbol{x}_t,\boldsymbol{u}_t) )^2\right]
\end{equation}

The idea is to minimize the temporal difference error between the Q-value estimates and the backed-up estimates from the next state and reward, as this approximates the Bellman optimality condition.

While DQN has been successful in discrete action spaces, extending these methods to continuous action spaces presents challenges. Two notable non-actor approaches for continuous action spaces are Continuous Action Q-Learning (CAQL) \cite{ryu2020caqlcontinuousactionqlearning} and Constrained Continuous Action Q-Learning (cCAQL) \cite{cCAQL}. CAQL adapts the Q-learning framework to continuous actions by using a neural network to represent the Q-function and optimizing it with respect to actions. cCAQL improves upon CAQL by introducing constraints to the action selection process, which helps to stabilize learning and improve robustness in continuous action spaces. As an alternative to directly optimizing Q-functions, actor-critic algorithms have also been employed to extend deep Q-networks methods to continuous action-space problems by including an actor-network that approximates the action taken by maximizing the Q-function. These modern RL algorithms include TD3 \cite{fujimoto2018addressing} and Soft-Actor Critic (SAC) \cite{haarnoja2018soft}.

\subsection{Evolutionary Strategies in Reinforcement Learning}

Within policy optimization methods, in addition to algorithms that leverage policy gradients, it is also possible to use evolutionary algorithms. Both types of algorithms update the parameters to optimize the policy which takes states as inputs and outputs (optimal) control actions (Figure \ref{fig:deep-pi}). This distinction is not unlike evolutionary and gradient-based algorithms in traditional optimization problems, i.e., evolutionary algorithms simply provide an alternative framework for learning the policy parameters.
Evolutionary strategies (ES) are a class of data-driven optimization algorithms inspired by principles of biological evolution. These algorithms optimize policies by iteratively generating populations of candidate solutions, evaluating their fitness (performance), and selectively propagating the fittest individuals to subsequent generations through processes similar to mutation, recombination, and selection.

In the context of RL, ES can be used to optimize the parameters $\boldsymbol{\theta}$ of a policy $\pi_{\boldsymbol{\theta}}$ directly, without relying on gradient information. These policies are evaluated in the environment on an episodic basis with their cumulative return as shown by a parameterized variant of Equation \ref{eq:ret}:
\begin{equation}
    J(\boldsymbol{\theta}) = \mathbb{E}_{\boldsymbol{\theta}} \left[\sum^T_{t = 0}\gamma^tr_t(\boldsymbol{x}_t,\pi_{\boldsymbol{\theta}}(\boldsymbol{x}_t))\right]
\end{equation}
Instead of using the approximation of the Q-function (Equation \ref{eq:Q}) or using a return gradient estimate to optimise the policy parameters. ES-RL algorithms use the estimate of the $J(\boldsymbol{\theta})$ and directly update the parameters towards those policies that produce higher returns to improve performance \cite{JMLR:v15:wierstra14a}. The key advantages of ES-RL algorithms include the ability to be easily parallelized, making them computationally efficient for evaluating multiple candidate solutions simultaneously. Additionally, ES-RL algorithms are less susceptible to getting trapped in local optima compared to gradient-based methods, as they explore the parameter space more ``globally'' through population-based search and they do not rely on stochastic estimates of gradients, which are also computationally expensive. Furthermore, ES-RL methods can be more robust to the inherent noisiness often associated with stochastic gradient descent (SGD) methods used in policy gradient approaches. Given these advantages, there has been some research interest in the ES-RL. \textcite{salimans2017evolution} applied developed an ES-RL algorithm and evaluated it on MuJoCo and Atari environments resulting in comparable performance to policy gradient methods such as TRPO \cite{10.5555/3045118.3045319}. \textcite{WU2023100073} used a hybrid strategy of derivative-free optimization techniques to solve an inventory management problem with improved performance over the policy gradient method Proximal Policy optimization (PPO) \cite{schulman2017proximal}. 

\subsection{PID Controllers}
The PID controller is a widely used feedback mechanism employed in industrial control systems~\cite{seborg2016process}. The discrete PID controller calculates an error value $e_t$ in discrete time as the difference between a desired setpoint and a measured process variable, and applies a correction based on three parameters: proportional ($K_P$), integral time constant ($\tau_i$), and derivative time constant ($\tau_d$). 
Note that other parameterizations of these degrees of freedom are possible. 
The proportional term applies a control action proportional to the current error, providing an immediate response to deviations from the setpoint. The integral term accumulates the error over time and applies a control action to eliminate steady-state errors. The derivative term considers the rate of change of the error and provides a dampening effect to prevent overshoot and oscillations.  The discrete position form of a single PID controller is defined as
\begin{equation}\label{eq:positionPID} 
u_t = K_p e_t + \frac{K_p}{\tau_i} \sum^t_{t=0}e_t + K_p \tau_d (e_t - e_{t-1})
\end{equation}
where $e_t = x^*_{i,t} - x_{i,t}$, is the setpoint error of state $i$ at timestep $t$, with the setpoint $x^*_{i,t}$ of state $i$ at timestep $t$.

Tuning the PID gains refers to finding values of the parameters $K_p$, $\tau_i$, and $\tau_d$ that result in good closed-loop performance (often measured by integrated squared error, etc.) and is crucial for achieving desired control performance. 
As a result, many popular tuning methodologies have been developed, including the Internal Model Control \cite{IMC, SIMC} and relay tuning \cite{relay}. The first of these methods is a model-based technique and the second excites the system, and uses the response to estimate the three PID parameters.

The time-invariant PID structure can achieve good performance on (approximately) linear systems. 
Historically, this condition was often sufficient, as processes are often operated around a known setpoint in an approximately linear region; however, more recent applications in control of nonlinear systems (e.g., transient, intensified, or cyclic processes) motivate more advanced control strategies. 
Given a nonlinear system the PID parameters will be dependent on the operating point, which motivates a gain scheduled approach. Gain scheduling involves designing multiple PID controllers for different operating regions and switching between them based on the current process conditions. In industrial applications, a common approach to gain scheduling is through the use of lookup tables, where the PID gains are pre-computed and stored for different operating conditions or setpoints \cite{LUT}. More recently, data-driven, model-free approaches to gain scheduling have gained traction as they are able to design the control directly from a single set of plant input and output data without the need for system identification \cite{CAMPI20021337}. Despite the advancements in PID control, challenges remain in terms of the manual effort required for controller tuning, and the limited performance in highly nonlinear and time-varying systems. These challenges motivate the integration of PID control with data-driven and learning-based approaches, such as reinforcement learning, to leverage the strengths of both paradigms. While more advanced control strategies like Model Predictive Control (MPC) exist, our focus on PID control is motivated by the ability to leverage existing infrastructure and well-established systems in industrial settings, providing a practical and widely applicable solution. Furthermore, most industrial applications of MPC utilize a lower-level PID as the regulatory controller hence, highlighting the prominence of PID control in chemical processes.

\newpage
\section{Methodology}\label{sec:method}
This section presents the proposed control-informed reinforcement learning (CIRL) framework, which integrates PID control structures into the policy architecture of deep RL agents. The methodology covers the CIRL agent design, policy optimization algorithm, and implementation details. 
\subsection{Control-Informed Reinforcement Learning (CIRL) Agent}
\begin{figure}
    \centering
    \includegraphics[clip, trim=3cm 10.2cm 24cm 3cm, width=0.75\textwidth]{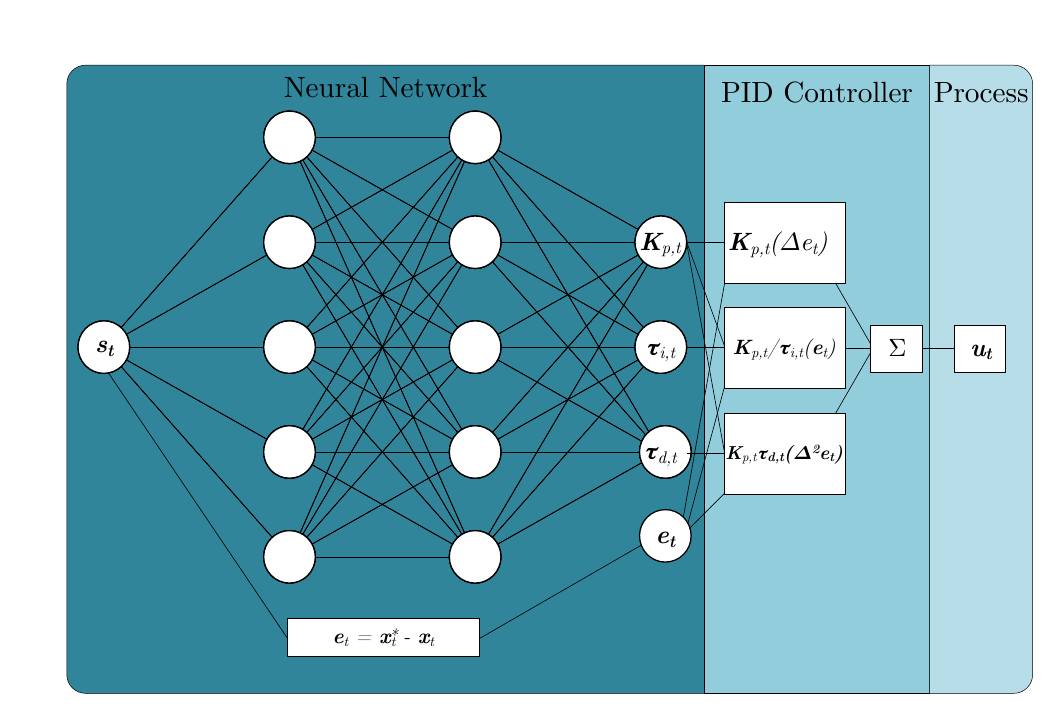}
    \caption{CIRL Agent}
    \label{fig:CIRL-agent}
\end{figure}

The CIRL agent consists of a deep neural network policy augmented with a PID controller layer, as illustrated in Figure \ref{fig:CIRL-agent}. The base neural network takes the observed states $\boldsymbol{s}_t$ as inputs and outputs the PID gain parameters $\boldsymbol{K}_{p,t}$, $\boldsymbol{\tau}_{i,t}$, and $\boldsymbol{\tau}_{d,t} \in \mathbb{R}^{n_u}$ at each timestep $t$. The PID controller layer then computes the control action $\boldsymbol{u}_t$ based on the error signal $\boldsymbol{e}_t = \boldsymbol{x}^*_t - \boldsymbol{x}_t$ and the current learned gain parameters.

The agent's state $s$ includes $N_t$ timesteps of history for both the state and setpoint, where $N_t > 2$ to fully define the velocity-form PID controller:
\begin{equation}
\boldsymbol{s}_t = \left[\boldsymbol{x}_{t\dots t-N_t}, \boldsymbol{x}^*_{t \dots t-N_t}\right]
\end{equation}

The PID layer is represented in the velocity form since, if the position form of the PID controller (Equation \ref{eq:positionPID}) is used and the gain changes suddenly, this can cause disturbances to the system \cite{velocityref}. The velocity form mitigates this issue by ensuring that the control input does not change abruptly despite sudden gain changes, and it is not necessary to reset the integral term. The $k^{th}$ PID controller of the system is represented by:
\begin{equation}
         \Delta u^{(k)}_{t} = K^{(k)}_{p,t} \Delta e^{(k)}_{t} + \frac{K^{(k)}_{p,t}}{\tau^{(k)}_{i,t}}e^{(k)}_{t} \Delta t  + K^{(k)}_{p,t}\tau^{(k)}_{d,t}\frac{\Delta^2 e^{(k)}_{t}}{\Delta t}
\end{equation}
where $\Delta e^{(k)}_t = e^{(k)}_t - e^{(k)}_{t-1}$, $\Delta^2 e^{(k)}_t = \Delta e^{(k)}_t - 2e^{(k)}_{t-1} + e^{(k)}_{t-2}$ and the superscript $(k)$ denotes the index of the controller, where $k \in {0, 1, \ldots, n_u}$, and $n_u$ is the total number of controllers in the system.

Through interacting with the environment, the CIRL agent aims to maximize the cumulative reward given by $r_t \in \mathbb{R}$ at each time step. For process control regulatory problems, various reward functions have been proposed. In general, they all involve some measure of (integrated) setpoint error, either squared or absolute, and/or a penalty for control action, similar to MPC objective functions. Adopting a similar notation to MPC, a squared error term penalizes deviations of the controlled variable from the setpoint, with larger deviations penalized more heavily: 
\begin{equation}
r_t = -\left(\boldsymbol{e_t}^T Q \boldsymbol{e_t} + \boldsymbol{u_t}^T R \boldsymbol{u}\right)
\end{equation}

where $\boldsymbol{Q} \in \mathbb{R}^{n_x \times n_x}$  and $\boldsymbol{R} \in \mathbb{R}^{n_u \times n_u}$ are weighting factors that balance the trade-off between tracking performance and control effort. 

It is important to note that derivative information is not passed between the PID controller and the neural network in the proposed CIRL agent architecture, as we take an evolutionary optimization strategy. Future work may study an integrated gradient-based learning strategy. The CIRL rollout pseudocode is given in Algorithm \ref{alg:CIRLRollout}
\begin{algorithm}[h!]
\caption{CIRL Rollout}
\label{alg:CIRLRollout}
\SetAlgoLined
\KwIn{Policy Parameters $\boldsymbol{\theta}$, Number of simulation timesteps $n_s$, Discrete time environment $f$}
\KwOut{Cumulative Reward $R$}
$\boldsymbol{s} \gets \boldsymbol{s}_0$ \tcp*{Reset observation to initial state}
$R \gets 0$ \tcp*{Initialize cumulative reward}

\For{$t = 0$ \KwTo $n_s - 1$}{
    ${K_{p,t}, \tau_{i,t}, \tau_{d,t}} \gets \pi_{\boldsymbol{\theta}}(\boldsymbol{s}_t)$ \tcp*{Get current PID gains from policy}
    $\boldsymbol{u}_t \gets \text{PID}(K_{p,t}, \tau_{i,t}, \tau_{d,t}, \boldsymbol{e}_t, \boldsymbol{e}_{t-1}, \boldsymbol{e}_{t-2})$ \tcp*{Use PID controller to output control input}
    $\boldsymbol{x}_{t+1}, r_t \gets f(\boldsymbol{u}_t, \boldsymbol{x}_t)$ \tcp*{Take one timestep in the environment}
    $\boldsymbol{s}_{t+1} \gets [\boldsymbol{x}_{t+1}, \boldsymbol{x}_{t}, \boldsymbol{x}^*_{t+1}]$ \tcp*{Update observation vector}
    $R \gets R + r_t$
}
\Return{$R$ \tcp*{Return cumulative reward}}
\end{algorithm}

\subsection{Implementation of Policy Optimization}
\begin{figure}
\centering
\begin{tikzpicture}[
    block/.style={rectangle, draw, text width=5cm, text centered, rounded corners, minimum height=1.2cm, font=\small},
    decision/.style={diamond, draw, text width=3.5cm, text badly centered, inner sep=0pt, font=\small},
    arrow/.style={->, >=stealth, thick},
    node distance=1cm
]

\node[block] (init) {Create and evaluate random initial policy parameters, select maximum};
\node[block] (update) at ($(init.south)+(0,-1cm)$) {Update particle velocities and positions};
\node[block] (evaluate_new) at ($(update.south)+(0,-1cm)$) {Evaluate new positions};
\node[block] (update_best) at ($(evaluate_new.south)+(0,-1cm)$) {Update personal and global best};
\node[decision, aspect=2] (check) at ($(update_best.south)+(0,-1.8cm)$) {$T$ iterations completed?};
\node[block] (output) at ($(check.south)+(0,-1.8cm)$) {Return optimal policy parameters};

\draw[arrow] (init) -- (update);
\draw[arrow] (update) -- (evaluate_new);
\draw[arrow] (evaluate_new) -- (update_best);
\draw[arrow] (update_best) -- (check);
\draw[arrow] (check) -- node[right, font=\small] {Yes} (output);
\draw[arrow] (check.west) -- ++(-1.5cm,0) |- node[left, pos=0.25, font=\small] {No} (update.west);
\end{tikzpicture}
\caption{Block diagram of the policy optimization algorithm}
\label{fig:policy_optimization}
\end{figure}
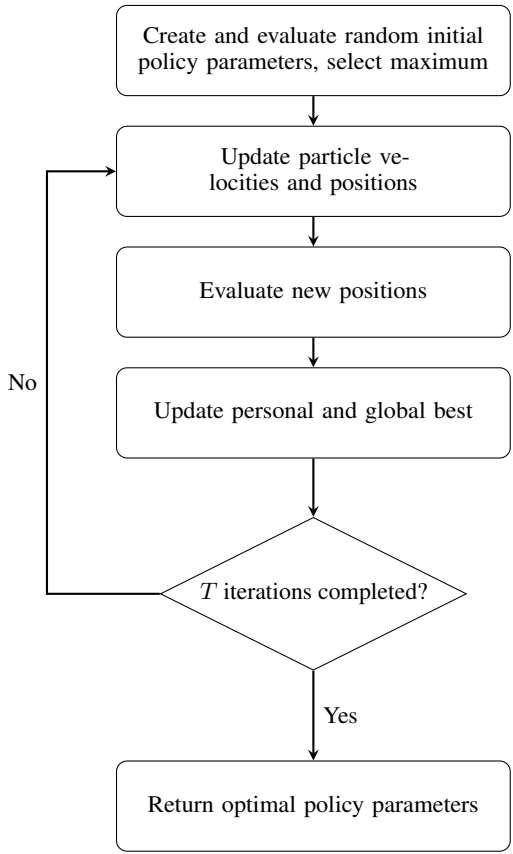

In this work, the CIRL agent's policy is optimized using a hybrid approach based on evolutionary strategies, combining random search and particle swarm optimization (PSO) \cite{PSOref}. A population of candidate policy parameter vectors is initialized by sampling randomly from the allowable ranges for each parameter dimension. The parameters in this case are the weights of the neural networks. This initial random sample provides a scattered set of starting points that encourages exploration of the full parameter landscape.
The random population undergoes $N$ iterations in which the objective function value (cumulative reward obtained by the policy in the environment) is evaluated for each policy to initialize the population in a good region of the policy space. The objective function to be maximized is:
\begin{align}
J(\boldsymbol{\theta}) = \mathbb{E}_{\pi_{\boldsymbol{\theta}}} \left[ \sum_{t=0}^{T} -(\boldsymbol{e_t}^T Q \boldsymbol{e_t} + \boldsymbol{u_t}^T R \boldsymbol{u})\right]
\end{align}
The best, or \textit{fittest}, policies from this initial random sampling are carried forward as seeds to initialize the PSO phase of the algorithm.
The PSO phase is then started, allowing the particles (policy parameter vectors $\boldsymbol{\theta}_i$) to explore areas around the initially fit random vectors in a more structured manner. In each PSO iteration, particle velocities and positions are updated as:
\begin{align}
\boldsymbol{v}_{i}^{t+1} &= w \boldsymbol{v}_{i}^{t} + c_{1} r_{1} \left( \boldsymbol{p}_{i}^{t} - \boldsymbol{\theta}_{i}^{t} \right) + c_{2} r_{2} \left( \boldsymbol{g}^{t} - \boldsymbol{\theta}_{i}^{t} \right) \label{eq:v_update} \\
\boldsymbol{\theta}_{i}^{t+1} &= \boldsymbol{\theta}_{i}^{t} + \boldsymbol{v}_{i}^{t+1} \label{eq:theta_update}
\end{align}

where $\boldsymbol{v}_{i}^{t}$ and $\boldsymbol{\theta}_{i}^{t}$ are the velocity and policy parameter vector of particle $i$ at iteration $t$, $w = $ is the inertia weight, $c_{1}$ and $c_{2}$ are the cognitive and social acceleration constants, $r_{1}$ and $r_{2}$ are random numbers in $[0, 1]$, $\boldsymbol{p}_{i}^{t}$ is the personal best policy parameter vector of particle $i$, and $\boldsymbol{g}^{t}$ is the global best policy parameter vector of the swarm.
This hybrid approach leverages the global exploration capabilities of initial random sampling, while also taking advantage of the PSO's ability to collaboratively focus its search around promising areas identified by the initial random search. The pseudocode for the policy optimization procedure is given shown by the block diagram (Figure \ref{fig:policy_optimization}). We highlight that the CIRL framework is agnostic to the policy optimization strategy, i.e., policy gradients or other policy optimization techniques can be used. Our choice of evolutionary algorithms in this work is motivated by optimization performance given the small size of the neural network, as well as robustness in training. 

\newpage
\section{Results and analysis} 
\label{sec:results}

\subsection{Computational Implementation}\label{sec:computational_implementation}

This section outlines the computational implementation employed in our study. We describe the state representation, neural network architecture, optimization parameters, and benchmark comparisons used to evaluate our proposed approach. Additionally, we provide information on the computational resources used. The RL state $s_t$ representation used for the CIRL agent included $N_t = 2$ timesteps of history, which is the minimum number of timesteps to define the PID controller layer:
\begin{equation}
\boldsymbol{s}_t = \left[\boldsymbol{x}_t, \boldsymbol{x}_{t-1},\boldsymbol{x}_{t-2}, \boldsymbol{x}^*_t, \boldsymbol{x}^*_{t-1},\boldsymbol{x}^*_{t-2}\right]
\end{equation}
The neural network architecture of the CIRL agent consists of three fully connected layers, each containing 16 neurons with ReLU activation functions, with the output being clamped to the normalised PID gain bounds. The CIRL agent is compared to a pure-RL implementation. This pure-RL agent consists solely of a deep neural network without the PID layer; we found this to require a larger network size and use three fully connected layers with 128 neurons.
While other architectures incorporating previous information, such as recurrent neural networks (e.g., LSTMs, GRUs), could be employed, we opted for this simpler structure in the present study. For the PSO algorithm used for policy optimization, we define the inertia weight $w$ as 0.6, while both the cognitive and social acceleration constants ($c_1$ and $c_2$, respectively) are set to 1. The policy optimization algorithm is initialized with $N = 30$ policies, before starting the PSO loop for $T = 150$ iterations with $n_p = 15$ particles. In this PSO loop, $n_e = 3$ episodes are used for each policy evaluation, with $n_s = 120$ timesteps in each rollout. All training was conducted on a 64-bit Windows laptop with an Intel i7-1355U CPU @ 3.7 GHz and an NVIDIA RTX A500 (Laptop) GPU. The CIRL agent required approximately 10 minutes of training time.

\subsection{CSTR Case Study}
\begin{figure}[h!]
    \centering
    \includegraphics[clip, trim=3cm 5cm 0.1cm 3cm, width=0.8\textwidth]{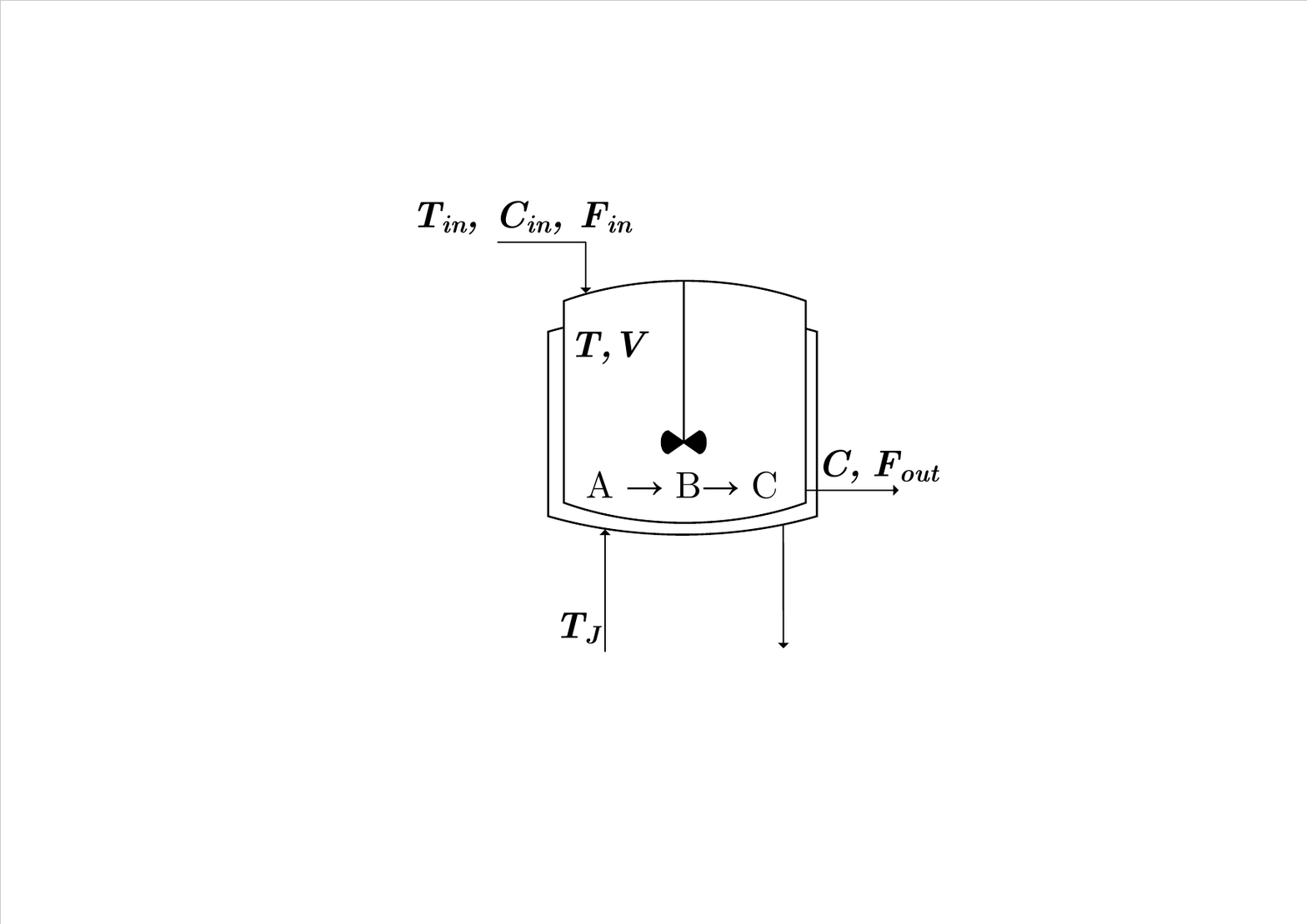}
    \caption{CSTR Process Flow Diagram}
    \label{fig:cstr_diagram}
\end{figure}

To demonstrate the proposed algorithm, simulation-based experiments were carried out on a CSTR system (Figure \ref{fig:cstr_diagram}) where both the volume and temperature are controlled. Though conceptually simple, this case study represents a non-trivial, multivariable system with nonlinear dynamics, capturing many challenges representative of those in real-world processes.  

The following generalized reactions take place in the reactor, where $B$ is the desired component:
\begin{equation}
A \underset{r_a}{\rightarrow} B \underset{r_b}{\rightarrow} C
\end{equation}
The following system of ordinary differential equations models the dynamics of the three chemical components in the reactor: $C_A$ (concentration of A in mol/m$^3$), $C_B$ (concentration of B in mol/m$^3$) and $C_C$ (concentration of C in mol/m$^3$), respectively.
\begin{align}
\dfrac{dC_A}{dt} &= \frac{F_{in}C_{A,in} - F_{out}C_A}{V} - r_a \\
\dfrac{dC_B}{dt} &= r_b - r_a - \frac{F_{out}C_B}{V} \\
\dfrac{dC_C}{dt} &= r_b - \frac{F_{out}C_C}{V}
\end{align}
where $F_{in}$ is the volumetric flow of feed into the system (m$^3$/min), $C_{A,in}$ is the feed concentration of species A (mol/m$^3$), $r_j$ is the reaction rate for reaction $j$ (mol/m$^3$/min), and $V$ is the volume of the CSTR (m$^3$). For this subsystem to be fully defined, the reaction rates are described by Arrhenius relationships for both reactions:
\begin{align}
r_a &= k_ae^{\frac{E_a}{RT}}C_A \\
r_b &= k_be^{\frac{E_b}{RT}}C_B
\end{align}
where $k_a, k_b$ are the Arrhenius rate constants (s$^{-1}$), $E_A, E_B$ are the activation energies (J/mol), $R$ is the universal gas constant (8.314 J/mol.K), and $T$ is the temperature (K).
The dynamics of the reactor temperature $T$ (K) and volume $V$ (m$^3$) are described by the following ordinary differential equations:
\begin{align}
\dfrac{dT}{dt} &= \frac{F_{in}(T_f - T)}{V}+ \frac{\Delta H_a}{\rho C_p} r_A + \frac{\Delta H_b}{\rho C_p} r_B + \frac{UA}{V\rho C_p} (T_c - T)\\
\dfrac{dV}{dt} &= F_{in} - F_{out}
\end{align}
where $T_f$ is the inlet stream temperature (K), $\Delta H_A, \Delta H_B$ are the heats of reaction (J/mol), $\rho$ is the density of the solvent (kg/L), $C_p$ is the heat capacity (J/kg/K), $U$ is the overall heat transfer coefficient (J/min/m$^2$/K), $A$ is the heat transfer area (m$^2$), and $T_c$ is the coolant temperature (K).
The parameters used in the simulation experiments are shown in Table \ref{tab:cstr_params}. The case study was implemented as a Gym environment~\cite{towers_gymnasium_2023} to provide a standardized format designed for RL algorithms. Within the gym environment, the system of ODEs are integrated using SciPy's ODEInt method.

\begin{table}[h!]
\centering

\caption{Parameters for the CSTR dynamic model}
    \begin{tabular}{@{}cc@{}}
        \toprule
        \textbf{Parameter} & \textbf{Value} \\
        \midrule
        $T_f$ & 350 K \\
        $C_{A,in}$ & 1 mol/m$^3$ \\
        $F_{out}$ & 100 m$^3$/sec \\
        $\rho$ & 1000 kg/m$^3$ \\
        $C_p$ & 0.239 J/kg-K\\
        $UA$ & $5 \times 10^4$ W/K \\
        $\Delta H_{a}$ & $5 \times 10^3$ J/mol  \\
        $E_{a}/R$ & 8750 K\\
        $k_{b}$ & $7.2 \times 10^{10}$ s$^{-1}$ \\
        $\Delta H_{b}$ & $4 \times 10^3$ J/mol\\
        $E_{b}/R$ & 10750 K \\
        $k_b$ & $8.2 \times 10^{10}$ s$^{-1}$ \\
        \bottomrule
        \label{tab:cstr_params}
    \end{tabular}
\end{table}

The three observed states of the reactor are the concentration of B $C_B$, reactor temperature $T$, and volume $V$, which define the state vector $\boldsymbol{x} = [C_B, T, V]$. We desire a policy that maps these to the action space, comprising the cooling jacket temperature $T_c$ and the inlet flow rate $F_{in}$, defining the control vector $\boldsymbol{u} = [T_j, F_{in}]$. This creates a system with two PID controllers, the first pairs $T_c$ and $C_B$ and the second pairs $F_{in}$ and $V$.The pairing was decided using a Relative Gain Array (RGA), which is shown in the appendix. This is additive measurement noise on all states of the CSTR. The system is simulated for 25 minutes with 120 timesteps. The bounds on the two control inputs are as follows $u^L$ = [290 K, 99 m$^3$/min] and $u^U$ =  [450 K, 105 m$^3$/min]. There are also bounds on the PID gains outputted by the DNN in the CIRL agent which are given in Table \ref{tab:PID_gains_bounds}. The initial state is defined as $x_0$ = [0 mol/m$^3$, 327 K, 102 m$^3$].

\begin{table}[h!]
\centering
\caption{Bounds on PID gains}
\label{tab:PID_gains_bounds}
\begin{tabular}{@{}lll@{}}
\toprule
         & $C_b$-loop  & $V$-loop \\ \midrule
$K_p$    &     [-5 , 25] ($\mathrm{K \cdot m^3/mol}$)       &  [0 , 1] ($\mathrm{s^{-1}}$)          \\
$\tau_i$ &     [0 , 20]   ($s$)    &      [0 , 2]   ($s$)    \\
$\tau_d$ &     [0 , 10]    ($s$)   &       [0 , 1]   ($s$)   \\ \bottomrule
\end{tabular}
\end{table}

\subsection{Training}

The CIRL and pure-RL algorithms were trained on nine setpoints that span the operating space of the CSTR case study (Figure \ref{fig:training_region}). The operating space is defined for $C_B$ between 0.1 and 0.8 mol/m$^3$ whilst maintaining a constant volume of 100 $m^3$. This is with the aim to learn a generalized control policy for a wide range of $C_b$ setpoints. Practically, this was achieved by rollout the policy on the three sub-episodes (1-3 in Table \ref{tab:training}) then summing them to create a single reward signal.

\begin{table}[h!]
\centering
\caption{Training and Test Scenarios}
\label{tab:training}
\scalebox{1}{
\begin{tabular}{@{}ccc@{}}
\toprule
\multirow{2}{*}{Sub-Episode} & \multicolumn{2}{c}{Setpoint Schedule}                                                                      \\
                             & \multicolumn{1}{c}{$C_B$ {[}mol/m$^3${]}}                            & \multicolumn{1}{l}{$V$ {[}m$^3${]}} \\ \midrule
1                            & 0.1 $\rightarrow$ 0.25 $\rightarrow$  0.4  & 100                                 \\
2                            & 0.55 $\rightarrow$ 0.65 $\rightarrow$ 0.75 & 100                                 \\
3                            & 0.7 $\rightarrow$ 0.75 $\rightarrow$ 0.8   & 100                                 \\ 
Test &\textbf{0.075}$\rightarrow$  0.45 $\rightarrow$   0.75 & 100\\
\bottomrule
\end{tabular}}
\end{table}

\begin{figure}[h!]
    \centering
    \includegraphics[clip, trim=0cm 0cm 0cm 0cm, width=0.45\textwidth] {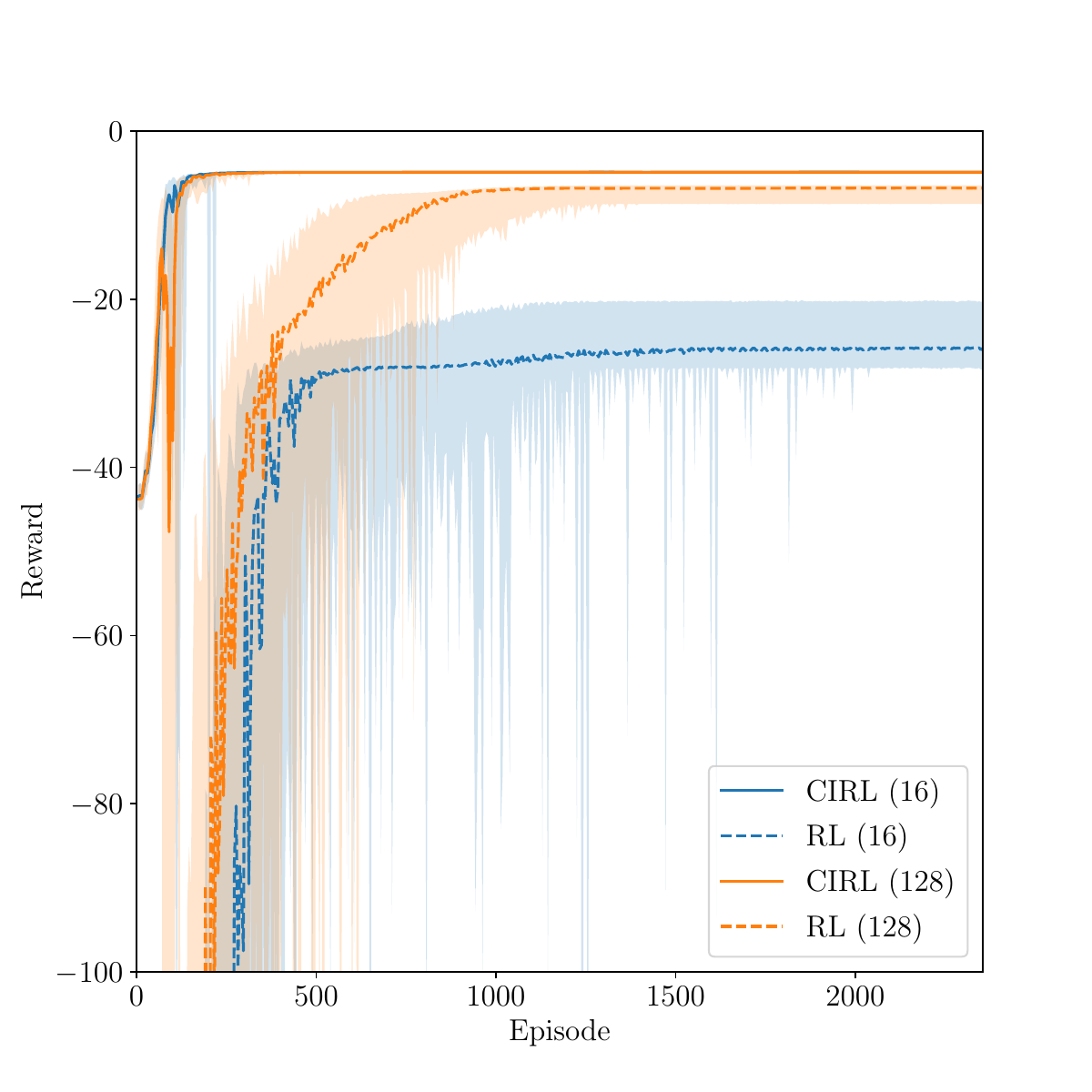}
    \caption{Learning curves for both RL and CIRL policies with 16 and 128 neurons per fully connected layer}
    \label{fig:network_size}
\end{figure}

As mentioned above, we found that, without the PID layer, a larger DNN policy was required to reach comparable performance. Therefore, the pure-RL algorithm implemented with a larger number of neurons (128) in each fully connected layer still reaches comparable training performance to CIRL (Figure \ref{fig:network_size}). 

The sample efficiency of RL algorithms is one of the main concerns with their implementation. Here, we demonstrate the improved sample efficiency of CIRL compared to an RL algorithm with the PID controller removed. The CIRL agent can be seen to initialize at a higher reward than the pure-RL implementation, since it has prior knowledge of the control strategy and benefits from the inherent stabilizing properties. This leads to more efficient and faster learning compared to pure-RL approaches, as the agent can make informed decisions and requires fewer samples to learn a good policy. 
In the real world, this corresponds to fewer simulations/experiments before an adequate control policy is obtained. Furthermore, given the stabilizing properties of the PID layer, this results in a safer policy, which inherently maintains setpoint tracking by utilizing the setpoint error. The PID controller's ability to continuously adjust based on the error between the desired setpoint and the current state provides a fundamental safety mechanism. This makes the overall policy more robust and less prone to dangerous deviations, especially during the early stages of learning when the neural network component might produce unreliable outputs.

\begin{figure}[h!]
    \centering
    \includegraphics[clip, trim=0cm 0cm 2cm 2cm, width=0.35\textwidth]{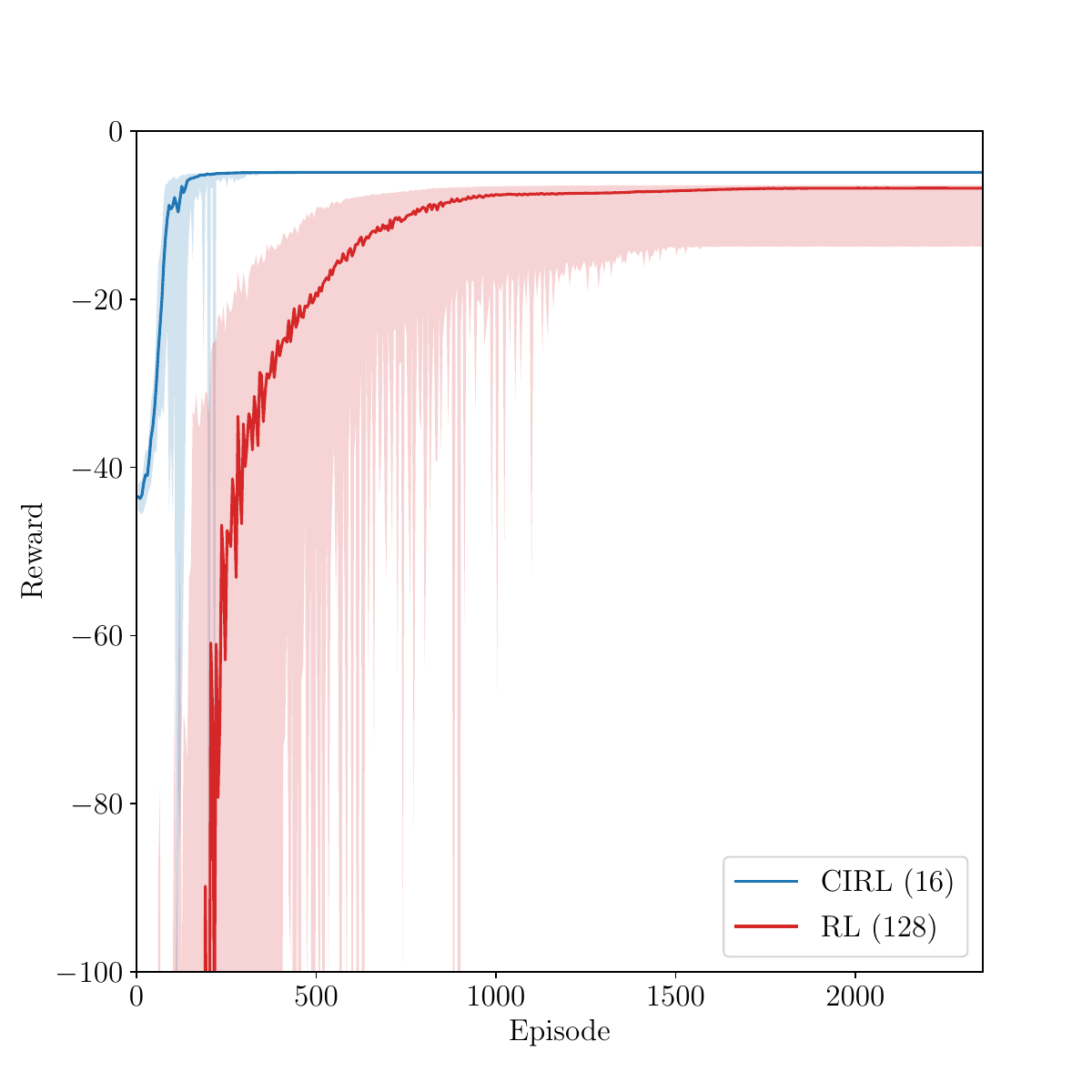}
    \caption{Setpoint tracking learning curves of CIRL and RL on 10 different seeds. Initial random search is omitted}
    \label{fig:learning_curve}
\end{figure}

The pure-RL agent, without any prior domain knowledge, needs to explore a larger number of samples, leading to slower convergence. This agent must learn the control strategy from scratch, including error correction and setpoint tracking that are inherently built into the CIRL approach. As a result, the pure-RL agent typically exhibits higher variance in its actions during the early stages of training, as it explores a wider range of potentially suboptimal strategies.
The lack of a PID layer means that the neural network in the pure-RL approach is learning to output controls which is inherently a larger space than the PID-gain space. This often necessitates a larger network architecture, as seen in our implementation with 128 neurons per layer, to capture the complexity of the control task. The increased network size, while providing more expressive power, also increases the dimensionality of the parameter space that must be optimized, potentially leading to longer training times and increased computational requirements. The learning curves over 75 iterations of the policy optimization algorithm for both the CIRL and pure-RL implementations are shown in Figure \ref{fig:learning_curve}.

\newpage
\subsection{Setpoint Tracking: Normal Operation}
\begin{figure}[h!]
    \centering
    \includegraphics[clip, trim=5cm 0cm 4cm 0cm,width = \textwidth]{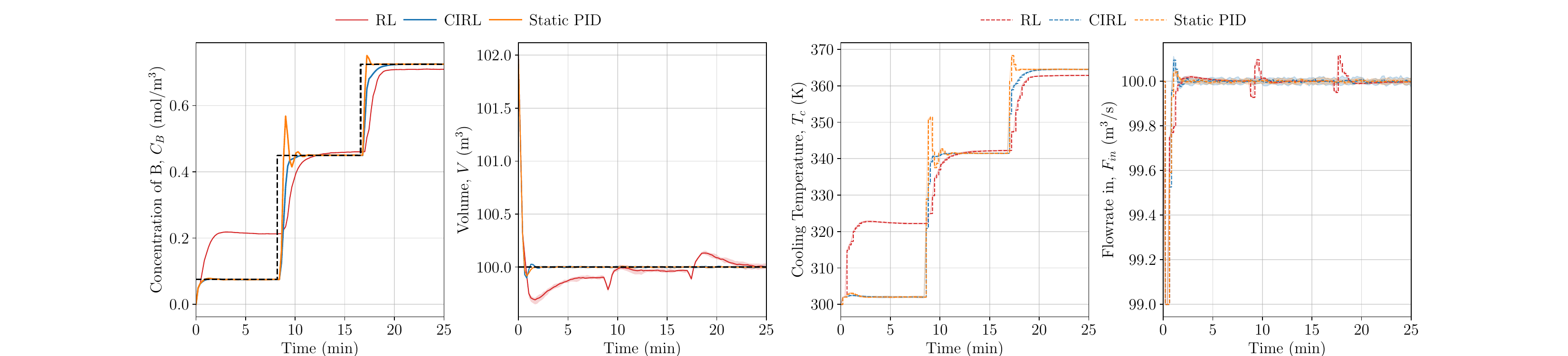}
    \caption{Setpoint tracking test scenario states and control inputs for CIRL, pure-RL and static PID}
    \label{fig:sp_track}
\end{figure}

We then test both learned policies on a partially unseen setpoint-tracking task. 
Specifically, the trained policies are then tested on a setpoint schedule detailed in Table \ref{tab:training}, which consists of three setpoints, the first setpoint is outside the training regime (shown in bold) and the other two interpolate between the training setpoints.  The CIRL agent is compared to the  pure-RL agent described previously in Section \ref{sec:computational_implementation} and a static PID controller. The static PID controller was tuned with differential evolution strategy to find gains using the setpoints in the training regime (Table \ref{tab:training}). The gains found for the static PID Controller are given in Table \ref{tab:pid_gains_static}. Then these three controllers were simulated on the test scenario (Table \ref{tab:training}) and shown in Figure \ref{fig:sp_track}.
\begin{table}[h!]
\centering
\caption{PID Gains for the Static PID Controller}
\label{tab:pid_gains_static}
\scalebox{1}{
\begin{tabular}{@{}lll@{}}
\toprule
         & $C_b$-loop & $F_{in}$-loop \\ \midrule
$K_p$    &     3.09 $\mathrm{K \cdot m^3/mol}$  &     0.84 $\mathrm{s^{-1}}$        \\
$\tau_i$ &     0.03 $s$      &     1.85  $s$       \\
$\tau_d$ &     0.83   $s$     &       0.08   $s$  \\ \bottomrule
\end{tabular}}
\end{table}

\begin{figure}[h!]
    \centering
    \includegraphics[clip, trim=1.5cm 0.01cm 1.3cm 0.01cm,width = 0.7\textwidth]{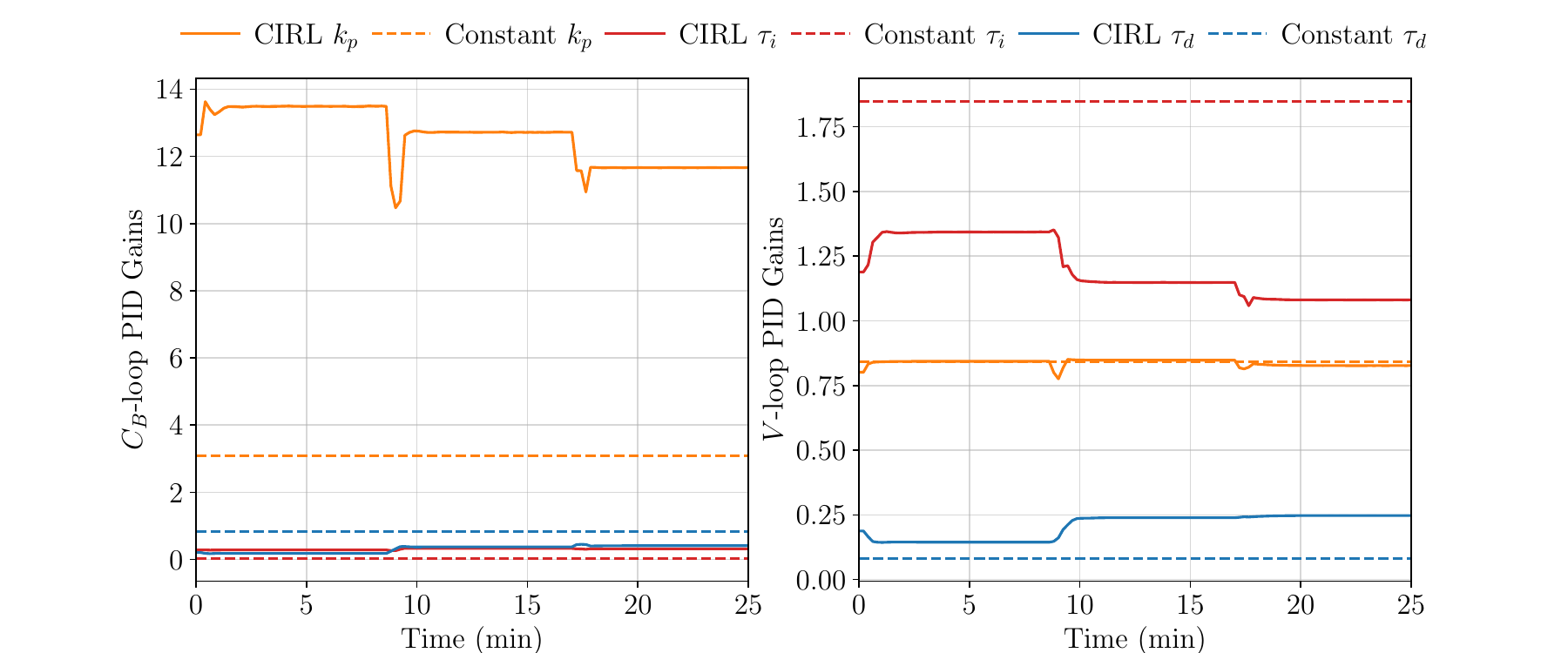}
    \caption{Gain trajectories for the $C_B$ and $V$ loop controllers}
    \label{fig:gains-sp}
\end{figure}

A conventional model-free implementation of deep RL (pure-RL in Figure \ref{fig:sp_track} exhibited poor tracking when generalising to these out-of-distribution setpoints ($x^*_{C_B} = 0.075$ mol/m$^3$) shown by the larger lower test reward (Table \ref{tab:test_reward}). By manipulating the proportional, integral, and derivative terms of its internal PID controller (Figure \ref{fig:gains-sp}), the CIRL policy could adapt its control outputs to track previously unseen setpoint trajectories. This ability to adaptively tune the PID gains allowed CIRL to outperform not only the model-free RL baseline approach, but also a static PID controller tuned to the setpoints in the training data. These results highlight the key benefit of the control-informed RL approach: integrating interpretable control structures like PID into deep RL enables performance gains compared to either component in isolation.

\begin{table}[h]
\centering
\caption{Final Test Reward for Pure-RL, CIRL and static PID}
\label{tab:test_reward}
\begin{tabular}{@{}cc@{}}
\toprule
Method        & Test Reward   \\ \midrule
RL            & -2.08          \\
\textbf{CIRL} & \textbf{-1.33} \\
Static PID    & -1.77          \\ \bottomrule
\end{tabular}
\end{table}
\newpage
\subsection{Setpoint Tracking: High Operating Point}
\begin{figure}[h!]
    \centering
    \includegraphics[width = 0.5\textwidth]{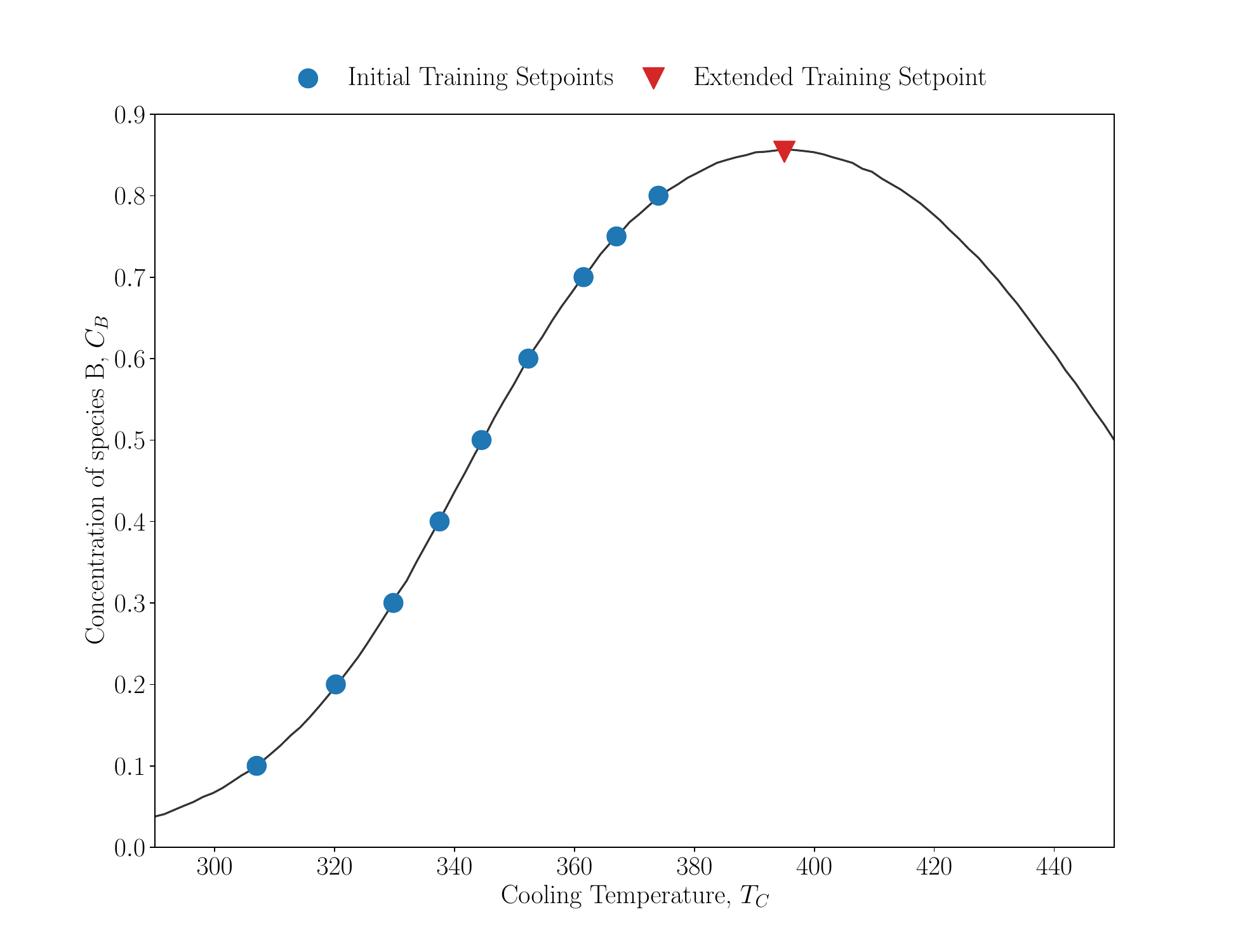}
    \caption{Operating space at a fixed $V$ = 100 m$^3$ with initial and extended training setpoints.}
    \label{fig:training_region}
\end{figure}

The CIRL algorithm does outperform the static controller in normal operation; however, the benefits are marginal and could potentially be attributed to an over-tuned controller. We now consider a more challenging operating scenario: if the operating point is pushed to a region of the operating space (red triangle in Figure \ref{fig:training_region}) the gradient changes significantly, as can be seen at cooling temperatures above 390 K. This is due to the second reaction rate increasing and consuming species B. This also poses a problem to the PID controller and PID layer in CIRL since to maximise the concentration of species B, the proportional gain must decrease and potentially change sign to stabilize around the maximum.

\begin{figure}[h!]
    \centering
    \includegraphics[width = \textwidth]{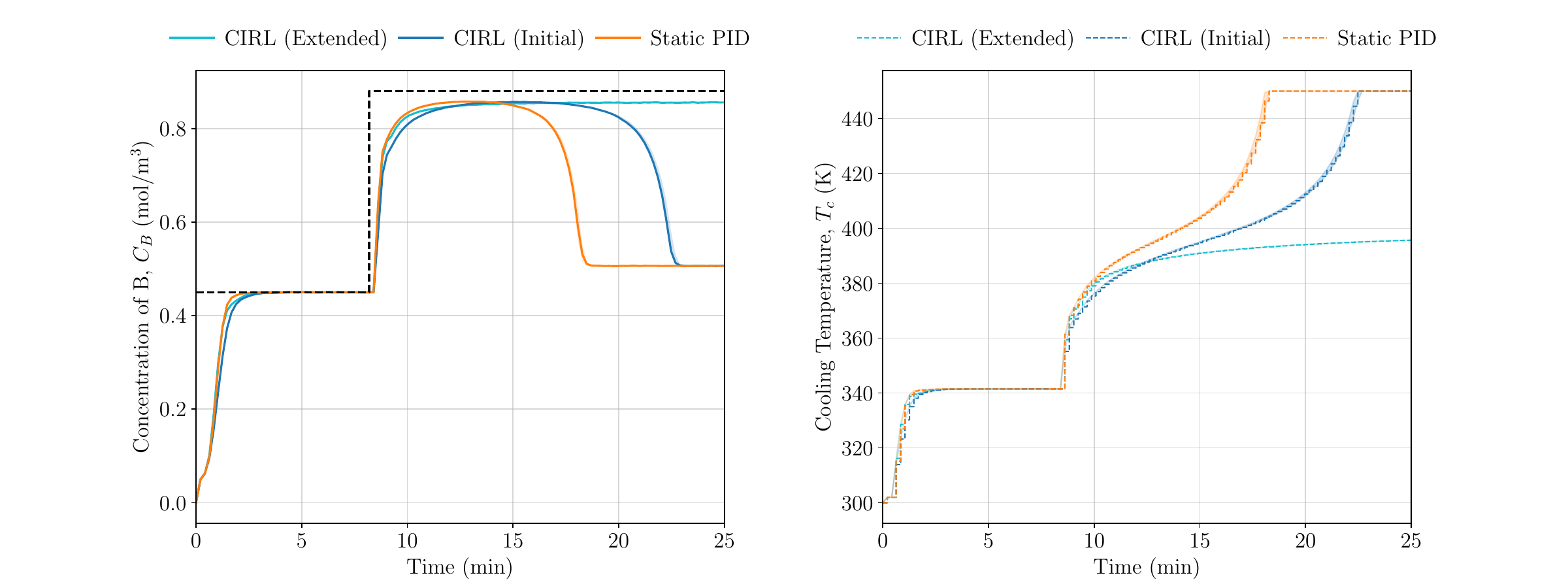}
    \caption{Setpoint tracking with the static PID control, initial, and extended training CIRL}
    \label{fig:sp_track_highop}
\end{figure}

\begin{figure}[h!]
    \centering
    \includegraphics[width = 0.5\textwidth]{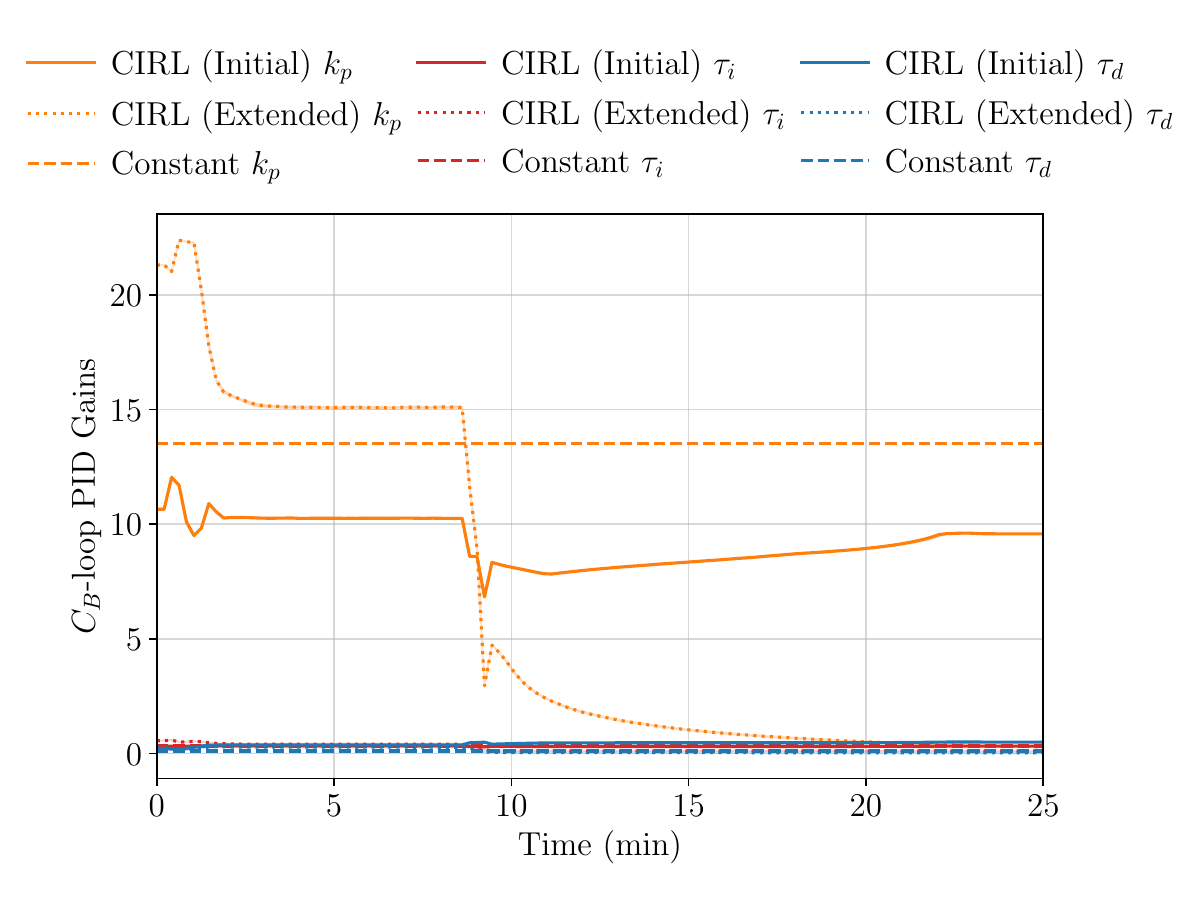}
    \caption{Gain Trajectories for the static PID control, initial, and extended training CIRL }
    \label{fig:gain_traj_highop}
\end{figure}

\begin{table}[h]
\centering
\caption{Final Test Reward for Pure-RL, CIRL and static PID}
\label{tab:test_reward_highop}
\begin{tabular}{@{}cc@{}}
\toprule
Method        & Test Reward   \\ \midrule
CIRL (initial)            & -4.04       \\
\textbf{CIRL (Extended)} & \textbf{-2.07} \\
Static PID    & -6.81          \\ \bottomrule
\end{tabular}
\end{table}
This scenario is explored by testing on a new setpoint schedule for species B of  0.45 to 0.88 mol/m$^3$ (Figure \ref{fig:sp_track_highop}). This high operating point scenario shows both the initial CIRL agent, as trained with a schedule shown in Table \ref{tab:training}, and the static PID controller both enter a closed-loop unstable regime since their gains remain at a large positive value. To attempt to negate this problem, the CIRL agent is trained on an extended training regime which includes the maximum of the operating region. This agent with an extended training regime decreases the proportional gain (CIRL Extended in Figure \ref{fig:gain_traj_highop}) which stabilizes the response of the controller.   

The scenario at high operating points reveals a limitation of the initial CIRL agent, as it enters an unstable closed-loop regime similar to the static PID controller due to the significant changes in gradient at cooling temperatures above 390 K. Nevertheless, the adaptability of the deep RL component of the CIRL framework is demonstrated by extending the training regime to include the upper limits of the operating space, allowing the agent to learn and adjust its control strategy, particularly by reducing the proportional gain, to maintain stability and achieve the desired setpoint even in the presence of these challenging conditions as shown by the higher test reward in Table \ref{tab:test_reward_highop}.

\subsection{Disturbance Rejection}
We now turn to evaluate the ability of the learned policies to reject disturbances. 
In particular, the CIRL algorithm is also tested on a scenario where there is a (unmeasured) step-change to the feed concentration of species A ($C_{A,in}$). Similar to the setpoint tracking case study, the CIRL algorithm is trained on multiple disturbance sub-episodes (Table \ref{tab:disttrain}). Then the trained agent is tested only on interpolation within this training regime. 

\begin{table}[h!]
\centering
\caption{Training and Test Scenarios}
\label{tab:disttrain}
\scalebox{0.9}{
\begin{tabular}{@{}ccc@{}}
\toprule
\multirow{2}{*}{Sub-Episode} & \multicolumn{1}{c}{Disturbance}                                                                                      \\
                             & \multicolumn{1}{c}{$C_{A,in}$ {[}mol/m$^3${]}}                            \\ \midrule
1                            & 1.5                                   \\
2                            & 1.6                                \\
3                            & 1.9 \\
Test                         & 1.75\\
\bottomrule
\end{tabular}}
\end{table}

\begin{figure}[h]
    \centering
    \includegraphics[clip, trim=0cm 0cm 0cm 0cm,width = 0.7\textwidth]{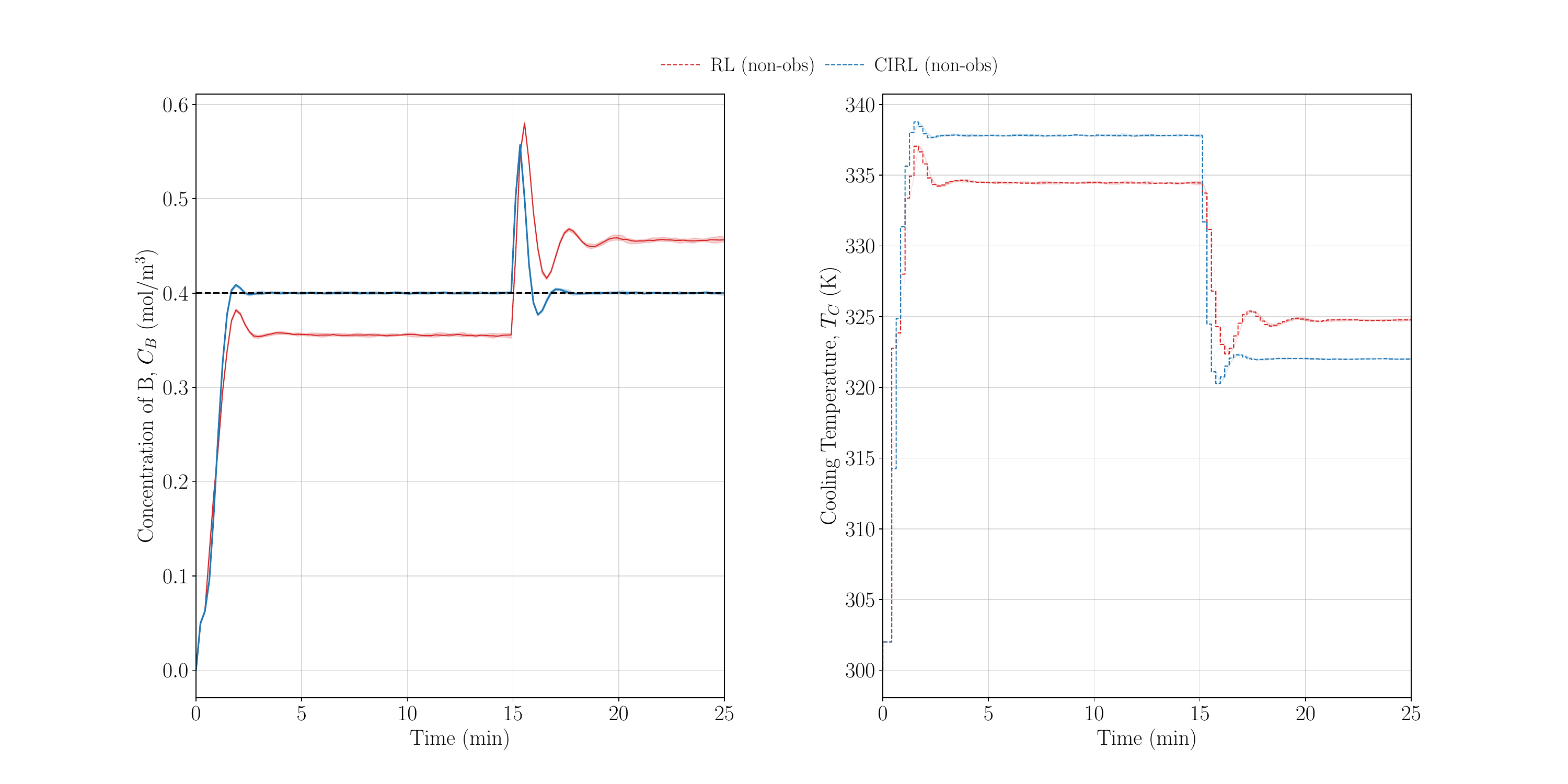}
    \caption{Disturbance rejection test scenario states and control inputs for CIRL, pure-RL with nonobservable disturbance}
    \label{fig:enter-label}\end{figure}

Under this disturbance condition, which effectively changes the underlying system dynamics, CIRL demonstrates a good ability to reject the disturbance and maintain the desired setpoint tracking performance. This disturbance rejection capability stems from CIRL's integrated PID control structure. The PID components in CIRL continuously measure and respond to the error between the setpoint and the actual system output, allowing it to adapt to and counteract unexpected disturbances in real-time, even if they were not explicitly modelled during training. Conversely, the pure-RL implementation exhibits poor setpoint tracking when faced with dynamics outside its training distribution (Table \ref{tab:test_reward_dist}). Without an explicit mechanism to handle disturbances, it settles for a compromised policy, i.e., sacrificing setpoint tracking performance both before and after the test disturbance occurred. This highlights that the addition of the PID components to the CIRL provide robustness to unmodeled disturbances. Unlike the pure-RL approach that attempts to anticipate and learn responses to all possible disturbances during training, CIRL's PID feedback mechanism allows it to adapt to unforeseen disturbances by using the measured error instead of modelling the response to the disturbance, demonstrating the fundamental advantage of closed-loop control in handling system uncertainties.

\begin{table}[h!]
\centering
\caption{Final Test Reward for Pure-RL, CIRL and static PID}
\label{tab:test_reward_dist}
\begin{tabular}{@{}cc@{}}
\toprule
Method        & Test Reward   \\ \midrule
CIRL            & -1.38       \\
pure-RL    & -1.76        \\ \bottomrule
\end{tabular}
\end{table}
\newpage
\section{Conclusion} \label{sec:conclusion}
This paper presents a control-informed reinforcement learning (CIRL) framework that integrates classical PID control structures into deep RL policies. A case study on a simulated CSTR demonstrates that CIRL outperforms both model-free deep RL and static PID controllers, particularly when tested on dynamics outside the training regime. The key advantage of CIRL lies in the embedded control structure, which allows for greater sample efficiency and generalizability. By incorporating the inductive biases of the PID controller layer, CIRL can learn effective control policies with fewer samples and adapt to novel scenarios more robustly than pure model-free RL approaches.

Future work may seek to incorporate additional existing information regarding existing PID infrastructure. 
For example, as a pre-processing step in the algorithm, the neural network could be initialized via offline reinforcement learning or behavioral cloning from past polices, potentially leveraging preexisting gain schedules in the plant. This initialization could potentially improve the starting point for the CIRL framework and accelerate learning.
Another direction may be enabling gradient-based training of CIRL agent by investigating the end-to-end differentiability of the PID controller layer.

The proposed CIRL framework opens up exciting research directions at the intersection of control theory and machine learning. This combination seeks to benefit from the best of both worlds, merging the known disturbance-rejection and setpoint-tracking capabilities of PID control with the generalization abilities of machine learning. Further investigations into theoretical guarantees, and online adaptation schemes have the potential to enhance the sample efficiency, generalization, and real-world deployability of deep RL algorithms for control applications across various industries.
\section{Acknowledgements}
Maximilian Bloor would like to acknowledge funding provided by the Engineering \& Physical Sciences Research Council, United Kingdom through grant code  EP/W524323/1. Calvin Tsay acknowledges support from a BASF/Royal Academy of Engineering Senior Research Fellowship
\section{Supplementary Information}
The code and data used within this work are available at \href{https://github.com/OptiMaL-PSE-Lab/CIRL}{{\fontfamily{qcr}\selectfont
https://github.com/OptiMaL-PSE-Lab/CIRL}}.
\newpage
\printbibliography
\newpage
\appendix
\section{Policy Optimization Algorithm}\label{appendix:alg}

\begin{algorithm}[h]
\caption{Policy optimization}
\label{alg:pol_opt}
\SetAlgoLined
 
\KwIn{
    $N$: number of initial policies, $n_p$: number of particles, $n_e$: number of episodes per evaluation, $n_s$: number of steps per episode, $T$: number of iterations}
\KwOut{Optimal policy parameters $\boldsymbol{\theta}^*$}

Initialize $\{\boldsymbol{\theta}_i\}_{i=1}^N$ randomly\tcp*{Initialize population}
\For{$i \in \{1, \ldots, N\}$}{
    $f_i \leftarrow 0$\tcp*{Initialize fitness}
    \For{$j \in \{1, \ldots, n_e\}$}{
        $R_j \leftarrow \text{CIRLRollout}(\boldsymbol{\theta}_i, n_s, f)$\tcp*{Run Algorithm 1: CIRL Rollout}
        $f_i \leftarrow f_i + R_j$\tcp*{Accumulate rewards}
    }
    $f_i \leftarrow f_i / n_e$\tcp*{Average fitness over episodes}
}
$\boldsymbol{x} \leftarrow \underset{\boldsymbol{\theta}_i}{\mathrm{argmax}} f_i$\tcp*{Select the best policy w.r.t. reward}
$\boldsymbol{g} \leftarrow \boldsymbol{x}$\tcp*{Initialize global best}
$\boldsymbol{p} \leftarrow \boldsymbol{x}$\tcp*{Initialize personal best position}
\For{$t \in \{1, \ldots, T\}$}{
    \For{$i \in \{1, \ldots, n_p\}$}{
        $\boldsymbol{v}_i^{t+1} \gets$ Eq. (\ref{eq:v_update})\tcp*{Update particle velocity}
        $\boldsymbol{x}_i^{t+1} \gets$ Eq. (\ref{eq:theta_update})\tcp*{Update particle position}
        $f_i^{t+1} \leftarrow 0$\tcp*{Initialize fitness for new position}
        \For{$j \in \{1, \ldots, n_e\}$}{
            $R_j \leftarrow \text{CIRLRollout}(\boldsymbol{x}_i^{t+1}, n_s, f)$\tcp*{Run Algorithm 1: CIRL Rollout}
            $f_i^{t+1} \leftarrow f_i^{t+1} + R_j$\tcp*{Accumulate rewards}
        }
        $f_i^{t+1} \leftarrow f_i^{t+1} / n_e$\tcp*{Average fitness over episodes}
        \If{$f_i^{t+1} > f(\boldsymbol{p}_i)$}{
            $\boldsymbol{p}_i \leftarrow \boldsymbol{x}_i^{t+1}$\tcp*{Update personal best if necessary}
        }
        \If{$f_i^{t+1} > f(\boldsymbol{g})$}{
            $\boldsymbol{g} \leftarrow \boldsymbol{x}_i^{t+1}$\tcp*{Update global best if necessary}
        }
    }
}
\Return{$\boldsymbol{\theta}^* = \boldsymbol{g}$}\tcp*{Return optimal policy parameters}
\end{algorithm}
\section{RGA Matrix}
The RGA matrix used for controller pairing is displayed below using averaged gains over three repetitions:
\begin{equation}  
RGA = 
\begin{bmatrix}
0.0003 & 0.9997 \\
0.9997 & 0.0003
\end{bmatrix}
\end{equation}
The values in the RGA matrix suggest a strong pairing between the first controlled variable ($C_B$ the concentration of B) and the second manipulated variable ($T_c$, the cooling temperature), and between the second controlled variable ($V$, the volume) and the first manipulated variable ($F_{in}$, the inlet flow rate).
\end{document}